\documentclass[preprint]{aastex}
\input epsf
\newcommand{\lref[1]}{\mbox{$L_{\rm ref}^{#1}$}}
\newcommand{\reff}{\mbox{$R_{\rm e}$}}
\newcommand{\mmax}{\mbox{$M_{\rm max}$}}

\newcommand{\msun}{\mbox{M$_{\odot}$}}
\newcommand{\lsun}{\mbox{L$_{\odot}$}}
\newcommand{\ishape}{{\tt ishape}}
\newcommand{\tinytim}{{\tt TinyTim}}

\newcommand{\bvz}{\mbox{($B\!-\!V)_0$}}
\newcommand{\vi}{\mbox{$V\!-\!I$}}

\newcommand{\ssfr}{\mbox{$\Sigma_{\rm SFR}$}}
\newcommand{\scl[1]}{\mbox{$\Sigma_{\rm cl}^{#1}$}}
\begin{document}
\title{The Luminosity Function of Star Clusters in Spiral Galaxies
 \footnote{Based on observations obtained with the NASA/ESA Hubble Space 
 Telescope, obtained at the Space Telescope Science Institute, which is 
 operated by the Association of Universities for Research in Astronomy, 
 Inc. under NASA contract No.\ NAS5-26555.}
}

\author{S{\o}ren S. Larsen
  \footnote{Guest User, Canadian Astronomy Data Centre, which
       is operated by the Herzberg Institute of Astrophysics,
       National Research Council of Canada}
  \affil{UC Observatories / Lick Observatory, University of California,
         Santa Cruz, CA 95064, USA}
  \email{soeren@ucolick.org}
}

\begin{abstract}
  Star clusters in 6 nearby spiral galaxies are examined using archive images 
from the Wide Field Planetary Camera 2 (WFPC2) on board the Hubble Space 
Telescope (HST).  The galaxies have previously been studied from the ground 
and some of them are known to possess rich populations of ``young massive 
clusters'' (YMCs). Comparison with the HST images indicates a success-rate 
of $\sim75$\% for the ground-based cluster detections, with typical 
contaminants being blends or loose groupings of several stars in crowded 
regions.  The luminosity functions (LFs) of cluster candidates identified on 
the HST images are analyzed and compared with existing data for the Milky Way 
and the LMC. The LFs are well approximated by power-laws of the
form $dN(L)/dL \propto L^\alpha$, with slopes in the range 
$-2.4\la\alpha\la -2.0$. The steeper slopes tend to be found among fits 
covering brighter magnitude intervals, although direct hints of a variation
in the LF slope with magnitude are seen only at low significance in two 
galaxies.  The surface density of star clusters at a reference magnitude of 
$M_V=-8$, \scl[-8], scales with the mean star formation rate per unit area, 
\ssfr .  Assuming that the LF can be generally expressed as 
$dN(L)/dL = c \, A \, \ssfr^\gamma \, L^\alpha$, where $A$ is the galaxy
area, $\gamma\sim 1.0 - 1.4$, $\alpha=-2.4$ and the normalization 
constant $c$ is determined from the WFPC2 data analyzed here, the maximum 
cluster luminosity expected in a galaxy from random sampling of the LF is 
estimated as a function of \ssfr\ and $A$.  The predictions agree well with 
existing observations of galaxies spanning a wide range of
\ssfr\ values, suggesting that sampling statistics play an important role in 
determining the maximum observed luminosities of young star clusters in 
galaxies.
\end{abstract}


\section{Introduction}

  Since the launch of the Hubble Space Telescope (HST), the ubiquity of 
highly luminous young star clusters in starburst environments, sometimes 
referred to as ``super star clusters'' (SSCs), has been firmly established. 
Some of the best-known cases include merger galaxies such as the 
``Antennae'' \citep{ws95} and NGC~3256 \citep{zepf99}, but SSCs have also
been found in galaxies that do \emph{not} show any obvious indications of 
having been involved in recent merger events.  Examples of isolated 
starburst dwarfs with SSCs are NGC~1569 and NGC~1705 \citep{as85,oc94}. The 
Large Magellanic Cloud also contains a number of 
``blue populous'' or ``young massive'' clusters (YMCs) \citep{sn51,hod61}, 
many of which have ages characteristic of Milky Way \emph{open} clusters, 
i.e.\ a few times $10^8$ years or younger, but are an order of magnitude more 
massive than any young cluster that is known in our Galaxy today.  Such 
clusters are also present in some spiral galaxies, 
and there appears to be a fairly tight relation between the star formation 
rates (SFRs) of galaxies and the number of YMCs 
\citep[][hereafter LR2000]{lr00}.

  The current terminology is confusing and reflects, to a large extent, a
lack of understanding of how different types of clusters are related.  The 
terms ``populous'', ``massive'' and ``super'' clusters were introduced as a 
reference to star clusters that apparently have no counterpart in our Galaxy, 
but do they actually constitute a separate class of objects, fundamentally
different from low-mass ``open'' clusters? Do they require special conditions 
to form? There is not really any \emph{a priori} reason to assume that the 
open cluster system in the Milky Way is representative for other galaxies or 
even other spirals, and by the same token, the \emph{presence} of YMCs in e.g.\ 
the LMC may be no more unusual than their \emph{absence} in our Galaxy. 
It may be more appropriate to view the cluster system of our Galaxy as part
of a continuum that ranges from very cluster-poor galaxies with low
SFRs such as IC~1613 \citep{wyd00}, over our Galaxy, to the LMC and finally
to starburst environments.

  Simple statistics may play a role in determining the luminosity of the 
brightest cluster in a galaxy, since random sampling of a power-law 
luminosity function (LF) will give brighter clusters in galaxies with richer 
cluster systems \citep{whit01}. Rich cluster systems, in turn, would generally 
be expected in galaxies with high levels of star formation.  In 
addition to statistical effects there may be a physical upper limit 
to the mass of star clusters that can form in a galaxy, determined by 
factors such as the gas density and -pressure of the interstellar medium 
which are also expected to correlate with the overall star formation rate 
\citep{ee97,ken98,bhe02}. It is currently unclear to what extent sampling 
statistics or physics dominate the maximum observed cluster luminosities 
and -masses, and under what conditions one or the other might prevail. 

  In reality, very little is known about how the LF of young cluster 
populations depends on environment.  \citet{vl84} found that the LF of 
Milky Way open clusters is well described as a power-law of the form 
$dN(L)/dL \propto L^\alpha$ with $\alpha=-1.5$ for $-8<M_V<-2$, but may 
steepen somewhere in the range $-11<M_V<-8$. Studies of young cluster 
populations in other galaxies generally find somewhat steeper slopes, 
$-2.5 \la \alpha \la -2.0$ \citep[e.g.][]{dk02,zepf99}, and \citet{whit99} 
suggested a bend at $M_V\sim-10.4$
for young clusters in the Antennae with $\alpha=-2.6\pm0.2$ for brighter
clusters and $\alpha=-1.7\pm0.2$ for fainter clusters.  Unfortunately, there 
is little overlap between observations of open clusters in the Milky Way, which 
cover a very limited section of our Galaxy and include few clusters 
\emph{brighter} than $M_V\approx-9$, and studies of young clusters in 
external galaxies where the identification of clusters \emph{fainter} than 
$M_V\approx-8$ is difficult because of the risk of confusion with individual, 
luminous stars. Even with HST, unambiguous identification of star clusters 
in crowded environments is problematic at distances greater than 10--15 Mpc.

  Considering the diversity of properties encountered among young cluster
populations even in normal spirals, HST images of nearby spirals may provide
some insight into correlations between properties of the cluster systems 
and host galaxy parameters.
  The HST archive contains a number of WFPC2 datasets covering nearby spiral
galaxies, including many of those studied by LR2000. At the typical
distances of these galaxies ($\approx 5$ Mpc), one WF pixel corresponds to
a linear scale of 2--3 pc, comparable to the typical half-light radius of
a star cluster.  This makes it possible to identify star clusters based on
their angular extent and thereby greatly reduce the risk of contamination by 
individual stars.  In this paper I examine HST archive images of 6 selected
galaxies from LR2000, some of which possess rich populations of YMCs while 
others have more modest, presumably more Milky Way-like cluster populations. 
As with any survey based on archive data, the choices of exposure times, 
filters and exact pointing are limited by what happens to be available. 
However, even with exposures in only one band one can perform interesting 
comparisons of the cluster populations in different galaxies. As a bonus, 
the WFPC2 data can be used for a check of the ground-based cluster 
identifications. The detailed structure of individual clusters will be 
investigated in a forthcoming paper.

\section{Data}
\label{sec:data}

  The data were downloaded from the HST archive at the Canadian Astronomy 
Data Centre (CADC) with standard ``on-the-fly'' pipeline processing. Only 
datasets including exposures longer than about 5 min in a $V$-band 
equivalent (F547M, F555W or F606W) were considered.  When several exposures 
in a band were available they were combined using the CRREJ task in the STSDAS 
package in IRAF\footnote{IRAF is distributed by the National Optical
Astronomical Observatories, which are operated by the Association of
Universities for Research in Astronomy, Inc.~under contract with the National
Science Foundation} to eliminate cosmic ray hits.  The WFPC2 datasets used 
in this paper are listed in Table~\ref{tab:datasets} and the pointings are 
shown in Fig.~\ref{fig:pointings}. As seen in Fig.~\ref{fig:pointings}, 
the pointings cover different parts of the respective galaxies relative to 
the center, spiral arms etc, and a more uniform coverage would certainly have 
been important in a dedicated survey. Nevertheless, the archive 
data used here should still provide a basis for a first rough intercomparison 
of star clusters in the different galaxies. In some cases the PC chip is
centered on the galaxy nucleus but the main focus in this paper is on 
clusters in the disks, and the nuclear regions, where crowding and dust
extinction may pose more serious problems, have generally been avoided.

  Objects were detected using the DAOFIND task in DAOPHOT \citep{stet87} 
running within IRAF, using a 3$\sigma$ detection threshold. When exposures 
in more than one 
band were available, DAOFIND was run independently on each exposure and the 
resulting object lists were matched to reduce the number of spurious 
detections.  Aperture photometry was then done with the APPHOT task in 
DAOPHOT, using aperture radii of 3 and 5 pixels.  When 
only one band was available, calibration of the photometry to standard $V$ 
magnitudes was provided by the zero-points in the 
\emph{WFPC2 handbook}\footnote{available at the URL 
http://www.stsci.edu/instruments/wfpc2}.
Otherwise the transformations in \citet{hol95} were used. 
Aperture corrections were determined by convolving a synthetic 
point spread function (PSF) generated by \tinytim\ \citep{kri97} with various 
models for the cluster profiles and measuring the flux within apertures of 
various sizes. The results are listed in Table~\ref{tab:apc} for aperture 
radii of 3 and 20 pixels, relative to a reference aperture of 5 pixels
\citep{hol95}.  In addition to point sources (the pure \tinytim\ 
PSF), the aperture corrections were determined for King and Moffat models 
\citep{lar99} with FWHM of 0.5 and 1.0 pixels, covering the typical range of 
intrinsic cluster sizes expected in the galaxies. The calibration by 
\citet{hol95} includes an implicit correction of 0.1 mag from their $0\farcs5$
reference aperture to infinity, which agrees fairly well with the 
aperture correction from $r=5$ to $r=20$ pixels for point sources
in Table~\ref{tab:apc}. 
However, magnitudes measured in the $r=5$ pixels aperture will be 
underestimated by an additional $\sim0.1$ mag for the most extended objects.  
For an aperture radius of 3 pixels, the magnitudes of objects with FWHM=0.5 
and 1.0 pixels will be too faint by about 0.12 mag and 0.35 mag, respectively 
(assuming King $c=30$ profiles), if the $-0.06$ mag aperture corrections for 
point sources are used.  For accurate integrated photometry of the clusters, 
knowledge of their intrinsic light profiles would 
be needed, but for the present purpose it was decided to simply use an
aperture correction of $-0.30$ mag for photometry in the $r=3$ pixels 
aperture.  Finally, note that the size-dependent corrections for \emph{colors} 
are generally negligible. 

  Sizes were measured for all objects using the \ishape\ code.  The 
code is described in detail in \citet{lar99}, and extensive tests of its 
performance have been carried out in that paper and elsewhere 
\citep[e.g.][]{lfb01}. Briefly, intrinsic object sizes are measured by adopting
an analytic model of the source and convolving the model with the PSF,
adjusting the shape parameters (FWHM, orientation, ellipticity) until the 
best match to the observed image is obtained. For the analytic model a
King $c=30$ profile was used, while the PSF was modeled with \tinytim .
When exposures of adequate depth were available in several bands, the sizes 
were obtained as an average of measurements in each band.

  Distances to the galaxies were taken from the references in \citet{lr99}, 
except for NGC~6946 where a distance modulus of 28.9 was assumed \citep{ksh00}.
Corrections for Galactic foreground extinction were applied according 
to \citet{sch98} as provided by the \emph{NASA/IPAC Extragalactic Database}
(NED). The adopted distance moduli and reddenings are listed in 
Table~\ref{tab:datasets}.  Even though the galaxies studied here are 
oriented nearly face-on so that the effect of internal extinction is 
minimized, significant extinction may still be present in some areas, 
especially within dusty regions in spiral
arms where $A_B$ can exceed 1 mag \citep[e.g.][]{kw01a,kw01b}. In 
Sect~\ref{sec:reddening}, $UBV$ data are used to obtain reddening estimates
for clusters in one of the fields, suggesting modest reddenings
for most of those clusters ($A_V \la 1$ mag). In the other fields there is 
no information available about the reddenings of individual clusters, 
so reddening within the galaxies has generally been ignored. Some
considerations concerning the effect of variable extinction on the observed 
luminosity functions are given in Sect~\ref{sec:masses}.

\section{Comparison of ground-based and WFPC2 data for clusters in NGC~6946}

  Because of the complete $UBVI$ coverage the data from HST Program 8715
(field NGC~6946--3) are very suitable for a comparison with the ground-based 
data used in LR2000.  
  WFPC2 images of all cluster candidates in this field identified from the 
ground are shown in Fig.~\ref{fig:cl6946f5} with each panel covering a 
$40\times40$ pixels ($4\arcsec\times4\arcsec$) section around each candidate.  
An extremely luminous cluster contained within the PC chip is discussed
elsewhere \citep{lar01} and has been omitted here.  Ground-based and WFPC2 
$UBVI$ photometry and sizes are listed in Table~\ref{tab:f5}.  Most of the 
objects in Fig.~\ref{fig:cl6946f5} are clearly extended and very likely 
clusters. With adequate S/N, objects with intrinsic FWHM down to 0.2 pixels
can be resolved by \ishape\ \citep[e.g.][]{lfb01}, corresponding to about 
0.8 pc at the assumed distance of NGC~6946. Only one object (\#2350) is smaller
than this limit, but this object is far too 
bright to be an individual star ($M_V=-10.8$, from the WFPC2 data) and is 
most likely a very compact cluster. For objects \#1490, \#2210 and \#2745 the 
classification as star clusters is more dubious, because these are located
in crowded areas where a blend of two or more stars might be mistaken by 
\ishape\ for a cluster. For this reason, these three objects have been
omitted from Table~\ref{tab:f5}. It therefore
appears that about 3 of the 13 objects in Fig.~\ref{fig:cl6946f5} might
be non-clusters, indicating a ``contamination'' fraction for the ground-based 
cluster identifications of $\approx 25\%$.  Inspection of ground-selected
cluster candidates on the full set of WFPC2 images in Table~\ref{tab:datasets}
corroborates this estimate and a similar conclusion was reached for a 
ground-based survey of young clusters in M51 \citep{lar00}.

  The WFPC2 photometry in Table~\ref{tab:f5} was done with an $r=5$ pixels 
aperture, correcting the $V$ magnitudes by $-0.1$ mag to account for the 
extended nature of the objects.  Even after that correction, the ground-based 
$V$ magnitudes remain somewhat brighter than those measured on the WFPC2 
images, with a mean difference of 
$\langle V_{\rm ground} - V_{\rm HST}\rangle = -0.37$ mag.
For \emph{colors} the systematic differences are very small, with 
$\langle (\ub)_{\rm ground} - (\ub)_{\rm HST}\rangle = 0.01$,
$\langle (\bv)_{\rm ground} - (\bv)_{\rm HST}\rangle = -0.03$ and
$\langle (\vi)_{\rm ground} - (\vi)_{\rm HST}\rangle = -0.02$.
The offset in $V$ magnitude is most likely due to the use of a much
larger aperture ($1\farcs5$) for the ground-based measurements, including
contributions from nearby stars and clusters which do not fall within
the smaller aperture used for the WFPC2 photometry.  To test if this can 
account for the difference between HST and ground-based photometry, $V$ 
magnitudes for the clusters were also measured in an $r=15$ pixels ($1\farcs5$) 
aperture on the HST images. These are listed in Table~\ref{tab:f5} as 
$V_{15}$, except for object \#1371 which is too close to the edge of the WF4 
chip. The measurements in this larger aperture do indeed agree very well with 
the ground-based ones, with a mean difference of only 
$\langle V_{\rm ground} - V_{\rm 15, HST}\rangle = -0.007$ mag.

  Concerning the sizes, it is clear that huge differences exist
between the measurements on ground-based and WFPC2 images. Again, this can
be attributed to the presence of other objects near the cluster candidates,
which tend to blur the ground-based images and make the clusters look
bigger than they are. This has been discussed in detail in
\citet{lar99phd}, where it was also shown that even relatively distant
``neighbors'' can affect the size estimates significantly. 

\section{Properties of Young Clusters in the WFPC2 fields}
  
\subsection{Selection of cluster candidates}

Even relatively short WFPC2 exposures allow cluster candidates to be 
identified with reasonable confidence to much fainter limiting 
magnitudes than from the ground because of the better angular resolution
and improved contrast.  Reliable size measurements require a 
S/N of 30--50 \citep{lar99}, which is reached in about 10 min with the 
F555W filter for a $V=23$ object superimposed on a moderately bright sky 
background, such as that encountered within the disk of a spiral 
galaxy. With the same exposure time, the broader F606W filter permits 
size measurements for somewhat fainter objects.

  Fig.~\ref{fig:v_sz} shows the intrinsic FWHM values as a function of 
$V$ magnitudes for all objects in the NGC~6946--3 field.
In this figure and throughout the rest of this paper, all photometry
was done in an $r=3$ pixels aperture.  For objects brighter than $V\sim23$, 
there is a fairly clear separation between extended sources and compact, 
unresolved objects which fall near the $x$--axis and have 
FWHM $\la 0.1$ pixel.
To reduce the risk of including contaminants in the cluster lists, relatively 
conservative size- and magnitude limits were chosen. For the
data in Fig.~\ref{fig:v_sz}, cluster candidates were selected as 
objects with $V<22.5$ and FWHM $> 0.2$ pixels. For the other fields the 
adopted magnitude cuts ranged between $V=22$ and $V=23$, depending on the 
quality of the data. To ensure homogeneity in the selection of 
cluster candidates, no other selection criteria, such as color cuts, were 
applied since some datasets contained exposures in only one band.

The magnitude cuts used for selection of
the cluster samples are well above the level where completeness 
effects are important. This is illustrated in Fig.~\ref{fig:cmpl}, which
shows completeness functions for each WFPC2 field. The completeness functions
were determined by adding artificial sources to each of the WF chips
at random positions (but with minimum separations of 10 pixels) and redoing 
the detection- and photometry procedures
and counting how many of the artificial sources that were recovered.
Objects with magnitudes between $V=20$ and $V=25$ were added at 0.5 mag
intervals, with 500 objects per chip at each magnitude step.
The artificial objects were modeled as King profiles with
an intrinsic FWHM of 0.5 pixels, convolved with the \tinytim\ point-spread 
function.  Note that the completeness is not a simple function of the exposure 
times (see Table~\ref{tab:datasets}), but depends on factors such as the sky 
background in the individual fields, degree of crowding etc.  The solid parts 
of the curves in Fig.~\ref{fig:cmpl} represent the magnitude intervals used 
for luminosity function analysis in Sec~\ref{sec:obslf}.  In all cases
the completeness is formally better than 93\% in the relevant magnitude 
intervals.  It should, however, be stressed that these tests only give a
rough indication of the actual cluster detection efficiency, since the 
clusters are not distributed at random within the frames but tend to clump 
together in ways that are not easily modeled. 

  The main concern is contamination of the cluster lists by other objects, 
of which blends of stars in crowded regions is the most serious problem.  
This effect is expected to be more severe at fainter magnitudes, possibly 
causing the slopes of the luminosity functions to be overestimated, and is 
considered below (Sect.~\ref{sec:obslf}).  Another potential problem is that 
some clusters might be more compact than the adopted size limit and therefore 
would not be counted as clusters, but to first order this should not lead to 
any systematic errors in LF slopes because the linear sizes of star clusters 
are nearly independent of their masses/luminosities 
\citep[e.g.][]{jan88,testi99,zepf99}.

  Background objects, especially early-type galaxies, might also resemble star 
clusters. To test how many such objects
would typically be expected in the relevant magnitude range within the WFPC2
field, F606W exposures of two reference fields were downloaded and reduced 
in the same way as the galaxy data. One field was the \emph{Hubble Deep Field}
\citep{wil96},
of which 7 exposures with a total exposure time of 6300 s were used, making
it much deeper than any of the galaxy datasets.  The other field was
a parallel WFPC2 exposure located 175\arcmin\ from NGC~628 (Program ID
9244), exposed for $500+260$ s in F606W and thus of about the same
depth as the galaxy datasets.  In the HDF and NGC~628 reference fields, a
total of 10 and 13 objects brighter than $V=23$ and 5 and 6 objects brighter 
than $V=22$ were found, of which only about half met the size criterion used 
for cluster selection. In the presence of dust extinction in the cluster host 
galaxies, these numbers would be further reduced.  The density of distant 
background galaxies may vary across the sky, but contamination by background 
galaxies is clearly not expected to pose any significant problem 
unless a rich galaxy cluster happens to be in the background. 

  Fig.~\ref{fig:bv_ub} shows a (\bv,\ub) two-color diagram for cluster
candidates in NGC~6946 with $UBV$ photometry (Table~\ref{tab:cl6946f5}), 
corrected for Galactic foreground reddening.  The solid curve superimposed 
on the plot represents the mean colors of LMC clusters \citep{gir95} while 
the dashed line is a 4 Myr solar metallicity stellar isochrone from 
\citet{gir02}.  There are a few objects with redder colors than expected, 
which could be highly reddened clusters or possibly background galaxies, 
but the majority of the cluster candidates have colors that are compatible 
with the LMC clusters.  Although the location of the bluest clusters in the 
(\bv,\ub) plane is not very different from that of individual early-type 
stars, the fact that most of the cluster candidates fall near the expected 
location provides a good sanity check on the identifications. 

The color distributions for cluster candidates in fields with photometry in 
more than one band are compared in Fig.~\ref{fig:colfig}. The information 
contained in a single color is limited, but slight differences between the 
fields do exist and are most likely due to different reddening and/or age 
distributions.  In particular, the broader \vi\ color distribution in the 
NGC~5236 field may suggest larger variations and a higher mean reddening 
in this field compared to NGC~6946--3.  However, multi-color data are needed 
to resolve the age-reddening degeneracy.

\subsection{Reddening}
\label{sec:reddening}

  For clusters with complete $UBV$ photometry, an estimate of the
individual reddenings can be obtained using the ``$Q$''-method 
\citep[hereafter VH68]{vh68},
where the intrinsic color (e.g. \bvz) of the cluster is assumed to be
a function of the reddening-free $Q$-parameter.  The reddening $E(\bv)$ is 
then $E(\bv) = (\bv)_{\rm obs} - \bvz$, and can be converted to e.g.\
$A_V$ using the reddening law in \citet{car89}, $A_V = 3.0 \, E(\bv)$.

  According to VH68, $Q = (\ub) - 0.75 \times (\bv)$, is related to the 
intrinsic color $\bvz$ of a cluster as $\bvz = 0.78\times Q + 0.40$.  
Comparison with population synthesis models (Bruzual \& Charlot 2001, 
priv.\ comm., hereafter BC2001) shows that a shallower relation, 
\begin{equation}
  \bvz \sim 0.24 \, Q \, + \, 0.21 
  \label{eq:q_bv}
\end{equation}
provides a better fit over most of the range in $Q$ (Fig.~\ref{fig:q_bv}). 
The range over which Eq.~\ref{eq:q_bv} provides a reasonable fit to the 
models is 10 Myr $\la$ age $\la$ 400 Myr, according to the BC2001 models.
It is, however,  important to note that the relation is strictly valid only for 
intermediate $Q$ values.  For $Q\la-0.65$, corresponding to ages below 
$\sim 30$ Myr, $\bv$ (or any other color) is no longer a simple function of 
$Q$ and the $Q$-method may underestimate $A_V$ by up to $\sim1$ mag for the 
youngest clusters. The situation is further complicated by the fact that
the shape of the ``loop'' in the colors of the youngest clusters is 
strongly metallicity-dependent.  A similar effect exists for $Q\ga-0.3$ 
(age $\ga$ 160 Myr), although the effect would here be to grossly 
overestimate the reddenings. 

  Figure~\ref{fig:ab} shows plots of $A_V$ vs.\ $Q$ parameter (a) and $V$ 
magnitude (b) for the clusters in Table~\ref{tab:cl6946f5}.  The horizontal 
dashed line indicates the Galactic foreground reddening towards NGC~6946, 
which is $A_B = 1.48$ mag or $A_V = 1.11$ mag according to \citet{sch98}. The 
plots do show some clusters with apparently very large reddenings, but for 
most clusters the observed reddenings are compatible with the Galactic 
foreground values and the internal absorption in NGC~6946 generally appears 
to be less than about 1 mag in $A_V$. Similar conclusions can be reached from
inspection of Figure~\ref{fig:bv_ub}.  For some clusters the derived 
reddenings are actually smaller than the foreground value, which might be 
partly due to the ambiguity in the reddening determinations for the youngest
clusters. Another uncertainty arises from stochastic color variations,
especially in low-mass clusters where the integrated light is 
dominated by a few individual stars \citep{gir95}. The results obtained
here do, of course, not necessarily give a representative picture of 
the situation in the other fields, but they are suggestive that most
clusters are subject only to modest reddenings. 

\subsection{Luminosity functions}
\label{sec:obslf}

  The LFs of cluster candidates in each galaxy are shown in 
Figs.~\ref{fig:lf1} and \ref{fig:lf2}. Solid line histograms represent the
raw, uncorrected LFs, whereas the dotted and dashed lines show the LFs after 
visual inspection of the cluster samples and removal of potential contaminants 
(see below).  Superimposed on each panel is a dotted-dashed line 
representing a power-law fit of the form
\begin{equation}
  dN(L_V)/dL_V = \beta \, L_V^{\alpha}
  \label{eq:lf}
\end{equation}
  to the raw, uncorrected luminosity distributions. In the following the 
subscript ``$V$'' will generally be omitted.  For comparison, data for LMC 
clusters taken from \citet{bica96} have also been included. According to
\citet{bica96}, their catalog is complete to about $V=13$ or $M_V \approx-6$,
assuming a distance modulus of 18.5.

  The data in Figs.~\ref{fig:lf1}--\ref{fig:lf2} are rather inhomogeneous in 
terms of absolute magnitude coverage, partly because the galaxies are at
different distances and partly because of the different exposure times.
The various fits have exponents between $\alpha\sim-2.0$ and $\alpha\sim-2.4$,  
with a tendency for the slopes to get steeper for cluster samples that span 
brighter magnitude intervals. However, a hint of a bend in the LF is seen 
directly in the data only for NGC~1313 and perhaps NGC~5236.  The lines drawn 
for these two galaxies represent fits to clusters with $-8<M_V<-6$.
The LF slopes are consistent with data for NGC~3627 presented by \citet{dk02}, 
who estimated $\alpha=-2.53\pm0.15$ for young clusters with $-11<M_V<-8$ in 
that galaxy.  \citet{ef85} found a shallower slope of $\alpha=-1.5$ for the 
LF of LMC clusters, based on eyeball estimates of the integrated magnitudes 
for a sample of 137 clusters, but the \citet{bica96} data indicate a slope 
of about $\alpha=-2$, compatible with the range of values found for the other 
galaxies studied here.  

  Table~\ref{tab:lffit} lists the parameters for the LF fits, given in two 
different formats. The magnitude intervals used for the fits and the number 
of clusters in each interval are given in columns (2) and (3).  The $a$ and 
$b$ values (columns 4 and 5) are the coefficients in a fit of the form
\begin{equation}
  \log \scl[]\ [{\rm kpc}^{-2} \Delta {\rm mag}^{-1}] = b \, + \, a \, M_V
  \label{eq:lfvdb}
\end{equation}
where $\scl[]$ is the surface density of clusters (clusters per 
magnitude bin per kpc$^{-2}$).  $\alpha$ and $\beta$ in Eq.~(\ref{eq:lf})
and $a$ and $b$ in Eq.~(\ref{eq:lfvdb}) are related as 
\begin{equation}
  \alpha = -(2.5 a + 1)
  \label{eq:a_alpha}
\end{equation}
and
\begin{equation}
  \beta = - \frac{2.5}{\ln 10} 10^{b + 4.8 a}
  \label{eq:b_beta}
\end{equation}
for $L_V$ in units of $L_{\odot}$ and $M_{V,\odot} = 4.8$.  For NGC~1313 
and NGC~5236 there are two entries for fits in the range $-8<M_V<-6$ and 
$-9<M_V<-7.5$, respectively.  These fits again suggest a steepening at
brighter magnitudes, although this is formally detected only at the 
$\sim1 \, \sigma$ level in NGC~1313 and at even lower significance in 
NGC~5236.  

Column (7) gives the surface densities of clusters at 
$M_V=-8$ ($\scl[-8]$), based on the LF fits.  An absolute 
magnitude of $M_V=-8$ is useful as a 
fiducial point for comparison of the cluster densities because it allows Milky 
Way and LMC data to be compared with data for more distant galaxies.
  The cluster densities in Table~\ref{tab:lffit} were obtained simply by 
normalizing number counts to the WFPC2 field of view, but because the pointings 
cover different sections of their host galaxies, it should 
be kept in mind that these cluster densities are not necessarily 
representative for each galaxy as a whole.  Significant local variations are
present even within the WFPC2 frames.  The data for the Milky 
Way open cluster luminosity function are taken from \citet{vl84} and refer 
to the Solar neighborhood, while many of the WFPC2 pointings cover regions 
closer to the center of the galaxies. The \citet{bica96} catalog covers 
about $5\deg\times5\deg$ of the LMC.

   Comparison with the Milky Way data 
confirms the notion that the cluster system of our Galaxy is comparable to 
those of other spirals with modest star formation rates (NGC~628, NGC~3184 
and NGC~6744), but poorer than that of the LMC and NGC~6946 
(see Table~\ref{tab:lmax} for \ssfr\ values). It is also 
worth noting that the LMC cluster system by no means appears to be extreme. 
Although richer than that of the Milky Way, it falls short of 
galaxies like NGC~1313, NGC~5236 and NGC~6946.
It is interesting (though hardly surprising) to note that \ssfr\ and 
$\scl[-8]$ scale with each other. A conclusion as to whether or not the 
relation is linear is, however, not really warranted by the current data, but 
other studies suggest that the efficiency of cluster formation relative 
to field stars increases with \ssfr\ (LR2000). Specifically, 
\citet[][hereafter BHE02]{bhe02} found that the total $U$-band luminosity of 
all clusters $L(U)$ normalized to the galaxy area $A$ scales with \ssfr\ as 
$L(U)/A \sim \ssfr^{1.4}$.  

  As mentioned above, some of the objects included in the cluster lists
may be contaminants. To check how much this effect and completeness issues 
might influence the LF fits, the datasets were visually inspected and objects 
suspected 
to be non-clusters were manually removed from the cluster lists. As an example,
this resulted in the removal of 76 out of 370 objects in the three
NGC~6946 fields, which were typically located within crowded regions of the 
images and most likely blends of two or several stars. The dotted and
dashed lines in Figs.~\ref{fig:lf1} and \ref{fig:lf2} show the 
contamination-corrected LFs, with and without correction for incompleteness.
The incompleteness corrections have very little effect on the LF 
fits, and the vast majority of uncertain cluster identifications are in the 
faintest bins.  Some of the objects that were rejected during the visual 
inspection might be real clusters, so the dashed lines may be considered 
a pessimistic estimate of the contamination.  In most cases the LF fits were 
not strongly affected by the visual inspection and removal of suspected 
contaminants, with the $\alpha$--values typically becoming flatter by 
$\sim0.1$.  The contamination corrections might bring the slopes 
of some of the observed LFs closer to that of the LMC.  The changes in 
$\scl[-8]$ were also modest, generally resulting in a decrease of less 
than $\sim$25\% relative to the values in Table~\ref{tab:lffit}. One galaxy 
where the visual inspection caused significant depletion at the faint end 
of the LF is NGC~1313, but note that the limiting magnitude in this galaxy 
is fainter than in most of the other galaxies studied here. This, combined 
with significant crowding in the central pointing (where most cluster 
candidates are located), made the identification of many of the faintest 
sources in this galaxy ambiguous.

\section{Brightest clusters - physics or sampling statistics?}
\label{sec:statorphys}

\subsection{Relations between maximum cluster luminosities and galaxy
  properties}

  BHE02 suggested a relation between \ssfr\ and the maximum mass \mmax\
of a star cluster that can form in a galaxy, based on the correlation between
gas density  and star formation rate \citep{ken98} and the assumption of
pressure equilibrium between cluster-forming cloud cores and the ambient 
interstellar medium. The predicted relation is of the form
\begin{equation}
   \mmax \; \propto \, \ssfr^\eta
  \label{eq:cdens}
\end{equation}
where $\eta=2$ for constant cluster \emph{density} (the case considered
by BHE02) and $\eta=2/3$ for constant
cluster \emph{size}.  Although observations show that any correlation between 
linear sizes and masses of star clusters is weak, if present at all, 
it is unclear whether the same is true for the proto-cluster cloud cores. 
For example, \citet{az01} have suggested that the absence of a size-mass 
relation for clusters could be at least partly due to a higher star formation 
efficiency in more massive clouds, causing low-mass clusters to expand more 
than high-mass clusters after the gas is blown away. Depending on the 
importance of such effects, $\eta$ might have values anywhere between 
the two extremes of 2/3 and 2.

  If the mass of the most massive cluster is always regulated by physical 
processes in the interstellar medium, the exponent $\eta$ might in principle 
be determined by plotting \mmax\ versus \ssfr\ for a sample of galaxies.
However, if the total cluster population is too small then \mmax\ may 
never be realized and the maximum observed cluster mass observed in a 
galaxy will instead be the result of random sampling from the mass function.  
In most cases information about the masses of individual clusters is not 
available, at least not with sufficient accuracy, and one is instead 
restricted to studying luminosity- rather than 
mass distributions. Because only a small fraction of the most massive clusters 
in a galaxy with an extended star formation history
will also be very luminous, the luminosity of the brightest cluster in
a galaxy might be limited by sample statistics, even if the mass of the most
massive cluster is not.

  The aim of this section is to investigate under what circumstances the 
luminosity of the brightest cluster is determined by either physics or 
statistics.  Let us assume for now that there is a universal cluster 
luminosity function of the form (\ref{eq:lf}) and that
the normalization ($\beta$) is proportional to the area-normalized 
star formation rate \ssfr\ to some power $\gamma$, and the galaxy
area $A$, i.e.
\begin{equation}
  dN(L)/dL = c \, A \, \ssfr^\gamma (L/\lref[])^{\alpha}.
\end{equation}
  In practice it is convenient to normalize the LF at a reference 
luminosity, \lref[], rather than at 1 \lsun, since $c$ will otherwise depend 
strongly on $\alpha$.
  The luminosity of the brightest cluster $L_{\rm max,stat}$ expected from
random sampling of the LF can then be estimated as in BHE02 by requiring 
\begin{equation}
  c \, A \, \ssfr^\gamma \, \int_{L_{\rm max,stat}}^{\infty} (L/\lref[])^\alpha dL 
  \; = \; 1,
  \label{eq:norm}
\end{equation}
i.e.
\begin{equation}
  L_{\rm max,stat} = 
    \left[ - \frac{c}{\alpha+1} \, 
        \left(\frac{A}{{\rm kpc}^2}\right)
        \left(\frac{\ssfr}
	           {10^{-3} \, \msun \, {\rm yr}^{-1} \, {\rm kpc}^{-2}}
             \right)^\gamma
        \left(\frac{\lref[]}{\lsun}\right)^{-\alpha}
    \right]^{-1/(\alpha+1)} \lsun , \;
  \mbox{for } \alpha < -1.
  \label{eq:lmax}
\end{equation}

  Eq.~(\ref{eq:lmax}) allows the maximum cluster luminosity expected from 
random sampling of the LF to be estimated when \ssfr, $A$ and the normalization
constant $c$ are known.  The last column in Table~\ref{tab:lffit} lists $c$ 
values for $\alpha=-2.4$, $\gamma = 1.0$, 
$\lref[]=1.32\times10^5\,\lsun$ ($M_V=-8$) and the observed 
cluster surface densities at \lref[] .  The average is 
$c\sim7.4\times10^{-6}$, or $c\sim5.5\times10^{-6}$ for 
$\gamma = 1.4$. All galaxies except NGC~5236 fall well within a factor 
of 2 from this normalization, even though $\scl[-8]$ and \ssfr\ both
vary by an order of magnitude.  When calculating the normalization constants
it was ignored that the cluster densities in Table~\ref{tab:lffit} may 
be somewhat overestimated because of contamination, but because of the 
overall uncertain nature of the contamination and completeness corrections, 
it was decided to simply use the LF fits in Table~\ref{tab:lffit} at 
face value.

  Direct extrapolation of a power-law luminosity function with $\alpha=-2.4$ 
from the local surface density of clusters near the Sun at $M_V\approx-8$ 
\citep{vl84} indicates that the brightest cluster within $\approx 2$ kpc from 
the Sun should have $M_V\approx-9.6$.  This is quite compatible with the 
observations, since the brightest open clusters within this distance have 
$M_V\approx-10$ (e.g.\ $h$ and $\chi$ Per).  If the density of open clusters 
throughout the disk of our Galaxy is as high as in the Solar neighborhood 
then the brightest cluster expected in the Galaxy as a whole should have 
$M_V\approx-12$.

\subsection{Comparison with observations}
\label{sec:compobs}

  Table~\ref{tab:lmax} lists the surface area $A$, \ssfr\ and $M_V$ magnitude 
for the brightest cluster ($M_V^{\rm max}$) in a sample of nearby galaxies. 
Most of the galaxies are from LR2000 and BHE02, but data for 5 spirals 
recently observed at Lick Observatory have also been included. The latter 
have not previously been published, except for NGC~5194 \citep{lar00}, but 
the data reduction and cluster selection procedures closely followed the 
description in \citet{lar99}.  Clusters were identified on $UBV$ images 
taken with the Prime Focus Camera (PFCAM) on the Shane 3 m telescope, 
supplemented with H$\alpha$ data from the 1 m Nickel reflector to eliminate 
HII regions from the sample.  For some of the galaxies in the BHE02 sample 
the brightest cluster magnitudes are given in the $R$ rather than $V$ band. 
In these cases a $V-R$ color of 0.2 has been assumed, corresponding to a 
cluster age between $10^7$ and $10^8$ years (BC2001).  Galaxy areas are 
calculated using the $\log D0$ values given in the RC3 catalog and star 
formation rates are based on either IRAS far-infrared fluxes (LR2000, Lick 
data) or H$\alpha$ (BHE02). For the BHE02 galaxies, the galaxy areas have 
been corrected to account for the fact that only a fraction of the galaxies 
were searched for clusters.

  Figure~\ref{fig:sfr_vm} compares the various predictions for the maximum
cluster luminosities, based on galaxy areas and star formation rates,  with 
the data in Table~\ref{tab:lmax}.  Panels (a) and (b)
show the observed brightest cluster magnitudes $M_V^{\rm max}$ as function 
of $A \, \ssfr$ ($\gamma=1.0$) and $A \, \ssfr^{1.4}$ ($\gamma=1.4$) and 
straight lines representing the relation expected from random sampling of 
the LF (Eq.~\ref{eq:lmax}) for $\alpha=-2.4$. Panel (c) shows $M_V^{\rm max}$ 
vs.\ \ssfr\ and lines corresponding to $\ssfr^{2/3}$ and $\ssfr^2$. 
Panels (a) and (c) are similar to the two upper panels in Fig.~16
of BHE02, except that the normalizations of the random sampling predictions
are here based on the data in Table~\ref{tab:lffit} and the addition of the Lick
data. Also note that the normalization in panel (c) is arbitrary. Physically,
the normalization of this relation depends on the relation between
\ssfr\ and the pressure in cluster-forming cloud cores, and is not easily
quantified.  If the two outliers NGC~1569 and NGC~1705 are excluded, least 
squares fits to the datapoints in panels (a) and (b) in Fig.~\ref{fig:sfr_vm} 
yield slopes corresponding to
$\alpha = -2.55\pm0.14$ and $\alpha = -2.84\pm0.18$ for $\gamma=1.0$
and $\gamma=1.4$, respectively.  The standard deviations around the
fits are 0.85 mag and 0.89 mag.  In panel (c) the the best-fitting 
power-law is $L_{\rm max} \propto \Sigma_{\rm SFR}^{0.76\pm0.18}$ with a
scatter of 1.49 mag.  Thus, the slopes of the relations in panels (a) and (b) 
in Fig.~\ref{fig:sfr_vm} are consistent with random sampling from luminosity 
functions similar to those in Table~\ref{tab:lffit}, although the fits
formally suggest that the bright-end slopes may be somewhat steeper than the
$\alpha=-2.4$ value used here.

  It is, however, clear that a comparison based only on the slopes of the 
observed relations between galaxy properties and maximum cluster luminosities 
is unlikely to provide strong constraints on the mechanisms responsible for 
producing highly luminous clusters. The observed slopes are compatible with 
predictions based both on physical arguments and sampling statistics, but
the \emph{scatter} is much reduced when a dependency on $A$ is included. This 
suggests that sample statistics play an important role in determining the 
luminosity of the brightest clusters even in fairly large spirals. 

  While the slopes of the relations in Fig.~\ref{fig:sfr_vm}a,b agree closely
with the expectations, the observed maximum luminosities are systematically 
lower than the predictions.  For $\gamma=1.0$ and $\gamma=1.4$, the mean 
offset is 0.92 mag and 0.75 mag, excluding NGC~1569 and NGC~1705.  This 
difference may very well reflect uncertainties in the area normalization of 
the cluster surface densities, and/or the exponent $\alpha$, dust extinction 
etc.  Also, the LR2000 survey excluded clusters with H$\alpha$ emission and 
objects located in very crowded regions, introducing a bias against the very 
youngest and hence most luminous objects.  To remove the offset by changing
only the cluster densities would require the mean cluster densities within 
$D0$ to be a factor of 3 lower than the estimates in Table~\ref{tab:lffit}.  
The 
only spiral where the spatial distribution of (young) star clusters has been 
discussed in some detail is M31, where the clusters are strongly concentrated 
at intermediate distances from the center \citep{hod79}.  If the same is the 
case in the galaxies studied here then one can suspect from  
Fig.~\ref{fig:pointings} that the mean cluster densities are indeed somewhat 
lower than inferred from the WFPC2 pointings.  Alternatively, extrapolation 
of a power-law LF with $\alpha=-2.85$ from the observed $\scl[-8]$ would also 
remove the offset, although this would be incompatible with most of 
the fits in Table~\ref{tab:lffit}.  A more complete survey of clusters is 
necessary in order to establish the relation between \ssfr , luminosity
functions and the cluster densities with greater certainty. 

  As pointed out by BHE02 and \citet{whit01}, the two dwarfs NGC~1569 
and NGC~1705 both contain a few highly luminous clusters that are much 
brighter than expected from the general size of their cluster populations. It 
is interesting to note that these two galaxies fit better into the ``pure''
\ssfr\ --  $L_{\rm max}$ relation in panel (c). The luminosity function 
of star clusters in these galaxies might be peculiar and perhaps biased 
towards high-mass clusters.

\subsection{The scatter in the observed maximum luminosities}

  Although the observed behavior of $L_{\rm max}$ as a function of $A$ and
\ssfr\ is well accounted for by random sampling of a power-law LF in terms 
of the slope and, to a lesser extent, the normalization, some 
scatter remains in Fig.~\ref{fig:sfr_vm}.  Part of this scatter may be due 
to uncertainties in the distances ($d$) of the galaxies, but such 
uncertainties are partly compensated for by the fact that both the observed 
and predicted maximum cluster luminosities scale with distance: 
$L_{\rm max,obs} \propto d^2 \propto A$, while 
$L_{\rm max,stat} \propto A^{0.7} \propto d^{1.4}$ (for $\alpha=-2.4$).
Thus, even an error of a factor of two in the distance of a galaxy only 
leads to a deviation of about 0.5 mag in the 
$A\,\ssfr^\gamma - L_{\rm max}$ relation, substantially less than the 
observed scatter.  However, for the $\ssfr - L_{\rm max}$ relation there are 
no such mitigating circumstances to reduce the effect of distance errors 
since \ssfr\ itself is distance-independent.

  To test how much scatter would be expected if the maximum cluster luminosity 
is the result of random sampling, a set of Monte Carlo simulations were 
carried out.  In each simulation, clusters were picked at random from a 
population with a power-law luminosity distribution with $\alpha=-2.4$ until 
2000 clusters had been counted in the interval $-8.5 < M_V < -7.5$.  A 
histogram of the $M_V$ magnitudes of the \emph{brightest} cluster encountered 
in each of 500 such simulations is shown in Fig.~\ref{fig:lmaxmc}.  The mean 
and median of the distribution are $\langle M_V^{\rm max}\rangle = -14.1$ and 
Med($M_V^{\rm max}$) = $-14.0$, respectively, while direct integration
of the LF (as in Eq.~\ref{eq:norm}) yields $M_V^{\rm max} = -13.7$ for a 
cluster population of this size.  The standard deviation is $\sigma_V = 1.04$ 
mag, similar to the observed scatter in the upper panels of 
Fig.~\ref{fig:sfr_vm}.  If the high-luminosity tail is truncated at 
$M_V=-16$, e.g.\ by a physical upper limit to the mass spectrum, then the 
scatter would decrease to $\sigma_V = 0.81$ mag.

  Thus, we find that not only the slopes but also the observed scatter in 
panels (a) and (b) in Fig.~\ref{fig:sfr_vm} are fully accounted for by random 
sampling of the LF, and that the scatter increases significantly if the 
maximum cluster luminosity is plotted as a function of \ssfr\ only.  Several
lines of evidence therefore suggest that statistical effects related to sample 
size play an important role in determining the luminosities of the 
brightest clusters in galaxies.

\subsection{A note on masses}
\label{sec:masses}

  Physically, it would be more relevant to compare cluster \emph{mass} 
functions rather than LFs. However, useful estimates of cluster masses based 
on integrated photometry require fairly accurate knowledge of ages because 
mass-to-light ratios are very sensitive to age.  For a sample of clusters 
with a range of ages, the shape of the luminosity function will generally 
be different from that of the underlying mass function \citep{meu95,zf99}.

  If the role played by sampling statistics is as important as suggested
in section \ref{sec:compobs}, it may be difficult to find galaxies that
provide large enough samples of clusters to make detection of a physical 
upper limit to the LF possible. However, it is still possible that physical
conditions play a role in shaping the \emph{mass} distributions of cluster 
systems.  For a continuous star formation history, even relatively quiescent 
galaxies ought to form very massive clusters once in a while if only sampling 
effects were important. Such massive clusters should be able to survive
for long periods of time. The Milky Way does contain a number of rather
massive ``open'' clusters that are several Gyrs old \citep{fri95} and the 
2-Gyr old cluster NGC~1978 in the LMC \citep{fis92} is another example.  

\subsubsection{Simulated luminosity functions: The effect of age differences 
  and reddening}

  A simple illustration of how the luminosity function may differ from the
mass function is provided in Fig.~\ref{fig:lf_synt}, which shows a
simulated luminosity function for clusters with ages uniformly distributed
between $10^7$ and $10^9$ years. The mass function was assumed to be a 
power-law with exponent $-2$ for $10^3 < M < 10^5 \, \msun$. The $V$-band 
mass-to-light ratio was estimated as 
$M_V(1\msun) = 1.87 \log ({\rm age}) - 11.8$, obtained as a fit to
model calculations by BC2001.  The slope
of the composite luminosity function is clearly steeper than that of the
underlying mass distribution, as indicated by the dashed line which is
a fit to the interval $-11 < M_V < -8$ and has a slope of $\alpha=-2.72$.
If the mass distribution were a single, untruncated power-law then the 
combined luminosity function would, of course, be a power-law with the same
slope. The steeper slope of the LF in Fig.~\ref{fig:lf_synt} is due to the 
truncation of the mass function at $10^5\, \msun$, but more subtle effects 
like a change in the slope of the mass function at a certain mass would 
have similar consequences. 

  This type of simulations can also provide some insight into the effect of 
dust extinction on the observed LFs.  Two cases were considered: 1) random 
extinctions, varying between $A_V = 0$ and $A_V=1$ mag; 2) random extinctions, 
varying between $A_V = 0$ and $A_V = 9 - \log ({\rm age})$ for 
$7 < \log({\rm age}) < 9$, as above.  The latter case was
adopted as an approximation to a probably more realistic situation where 
the mean extinction decreases with age. With the extinctions added, the
clusters become 0.5 mag fainter on the average, so the fitting interval was
adjusted to $-10.5 < M_V < -7.5$ in order to probe the same mass range. For 
case (1), the slope is virtually unchanged at $\alpha=-2.69$, while case (2) 
gives an even steeper slope of $\alpha=-2.97$. This suggests that 
age-dependent extinction will steepen the LF slopes, but a correction 
for this effect would require detailed knowledge about the age distribution 
of the clusters involved and the variation of mean extinction with age.

\subsubsection{Masses for clusters in NGC~6946}

  $UBV$ data allow ages to be derived 
with reasonable accuracy and masses can then be estimated with help from 
population synthesis models.  Ages for all clusters in 
Table~\ref{tab:cl6946f5} were derived according to the ``S''-sequence 
calibration in \citet{gir95}, which is based on LMC clusters.  The metallicity 
of clusters in NGC~6946 may be different from that of LMC clusters, but 
metallicity has only a small influence on colors for young objects, except for 
clusters younger than about $10^7$ years where the integrated colors are 
strongly affected by the rapid, strongly metallicity-dependent and poorly 
understood evolution of red supergiants.  For such clusters, a simple relation 
between broad-band colors and age no longer exists and rather than quoting 
uncertain age estimates they are simply listed as $<10^7$ years in 
Table~\ref{tab:cl6946f5}.  Note that S-sequence ages are not strongly 
affected by reddening -- a $V$--band extinction of 1 mag will typically 
lead to the cluster ages being overestimated by $\sim0.2$ dex.  Finally, M/L 
ratios from the BC2001 population synthesis models were used to estimate 
the mass of each cluster.  The models used here assume a Salpeter stellar IMF
with a minimum mass of 0.1 \msun . More realistic IMFs show some
flattening below 0.5 \msun\ \citep[e.g.][]{kroupa01}, which will result
in a smaller total mass than for a Salpeter IMF extending to 0.1 \msun, 
but this does not affect the relative comparisons performed here.

  Fig.~\ref{fig:age_m} shows mass vs.\ age for the clusters 
listed in Table~\ref{tab:cl6946f5} and clearly demonstrates that any 
attempt to derive a meaningful mass distribution requires deeper photometry 
as well as a much larger sample of clusters.  Because of the rapid 
increase in M/L ratio as a function of age, illustrated by the dashed line,
only the most massive clusters are detectable at ages greater than $10^8$ 
years or so, and an appropriate mass range can be observed only for the 
youngest clusters. Essentially no clusters older than about 300 Myrs are 
detected.  Discarding a couple of clusters with very large and uncertain ages, 
the most massive clusters found in this single WFPC2 pointing have masses on 
the order of $2\times10^5 \, \msun$, comparable to the masses of the most 
\emph{luminous} clusters in NGC~6946 as a whole.  The observed mass 
distribution in this field is again most likely limited by sample statistics 
and offers little insight as to whether or not a real upper limit to the 
mass distribution of clusters exists in this galaxy.

\section{Summary}

  The cluster populations of 6 nearby spiral galaxies have been examined
on archive HST/WFPC2 images.  Reidentification of cluster candidates 
selected on ground-based images in previous work suggests a success-rate 
of roughly 75\% for ground-based cluster selection.  The ground-based 
photometry tends to overestimate the brightness of the clusters by up 
to several tenths of a magnitude, because other objects than the cluster
itself are included within the relatively large apertures used for the 
ground-based photometry. On the other hand, ground-based colors agree very 
well with the WFPC2 photometry.

  Typical slopes for the cluster luminosity functions 
$dN(L)/dL \propto L^{\alpha}$ studied here are
$-2.4 \la \alpha \la -2.0$, clearly steeper than the $\alpha=-1.5$ for the 
Milky Way open cluster LF found by \citet{vl84} but in good agreement with
data for LMC clusters taken from \citet{bica96}.  However, the Milky Way data 
cover a much fainter magnitude interval and van den Bergh \& Lafontaine
pointed out that the lack of very luminous clusters in the Milky Way by 
itself suggests a steepening of the LF at the bright end.  There is a general
tendency for fits covering brighter luminosity intervals to have steeper 
slopes, but a direct hint of a change in slope at $M_V\approx-8$ is seen 
only in two galaxies (NGC~1313 and NGC~5236), at the $\la1\sigma$ level.

  The surface densities of clusters ($\scl[-8]$) scale with the 
area-normalized star formation rates of the host galaxies.  Assuming that
this scaling is generally valid, the maximum cluster luminosity expected
from random sampling of the LF was estimated by extrapolation of a 
power-law LF with $\alpha=-2.4$, scaled to the area $A$ and \ssfr\ for
a sample of galaxies. Both the slope and scatter of the predicted relation
agree well with observations by \citet{lr00} and \citet{bhe02}, spanning a
wide range in \ssfr\ and physical dimensions.  The maximum cluster 
luminosities are, however, overpredicted by about 0.9 mag, possibly
because the mean density of clusters in the galaxies is lower than estimated
from the HST pointings.  Some starburst dwarfs do manage to form clusters 
that are significantly more luminous than expected from the size of their 
cluster populations, indicating that the LF of young clusters in these 
galaxies may be peculiar. 

  It remains to verified whether the cluster LF is really universal, 
concerning the slopes and their dependence on magnitude, as well as the 
location and even the existence of a ``bend'' in the LF. Data presented here 
suggest that a change in slope occurs at $M_V\sim-8$, 
but observations of a similar bend at $M_V\sim-10.4$ in the ``Antennae'' 
suggest a possible dependency on factors such as environment and/or the age 
distribution of the clusters involved.  An interesting point is whether or not 
the mass distribution observed in young cluster systems can be reconciled with 
that of old globular clusters (GCs). The low-mass end of the old GC mass 
spectrum is likely to be strongly affected by dynamical evolution 
\citep{fz01,ves00,ves01}, but for masses above $\sim10^5$ \msun\ the mass 
distribution is well approximated by a power-law with slope $\sim-1.9$ 
\citep{kis94,kis96,lar01b}.  This is somewhat shallower than the slopes of 
the luminosity functions at bright magnitudes found in this paper, but 
similar to the faint-end slopes.  The interpretation of these 
differences is complicated by the fact that the LF of a population of 
clusters with a range of ages is, in general, different from the more 
fundamental underlying \emph{mass} distribution.  In particular, if the
mass function has a physical upper limit then the LF of a cluster population
with a range of ages will tend to have a steeper slope than the underlying
mass distribution.  This is well illustrated by 
the finding that the mass function of Antennae clusters follows a single 
power-law with $\alpha=-2$ over the range 
$10^4 < M < 10^6\,\msun$ \citep{zf99}. 

  Because random sampling statistics appear to play an important role in 
determining the luminosity of the brightest clusters observed in galaxies, 
the role of physical processes in regulating the upper mass limit of star
clusters in galaxies is difficult to determine.
For galaxies with a continuous star formation history, 
observations of the cluster \emph{mass} distribution should make it possible 
to establish with greater certainty how the maximum possible cluster mass 
depends on 
host galaxy properties. This, however, would require fairly accurate age 
estimates for large samples of clusters, down to $M_V\sim-6$ or fainter.
The new Advanced Camera for Surveys on board HST will be very suitable for
such studies because of its large field of view and superior sampling
compared to WFPC2.  An essential ingredient in future surveys should be 
inclusion of $U$ band imaging, which would allow better age estimates as well 
as providing constraints on cluster reddenings. Multi-color photometry would
also make the cluster identifications more secure.

\acknowledgments

This work was supported by National Science Foundation grant number AST9900732
and by HST grants GO-08715.02-A and AR-09523.  This research has made use of 
the NASA/IPAC Extragalactic Database (NED) which is operated by the Jet 
Propulsion Laboratory, California Institute of Technology, under contract with 
the National Aeronautics and Space Administration.  I thank J.\ Brodie,
T.\ Richtler and the anonymous referee for helpful comments.

\newpage
\onecolumn

\clearpage
\noindent
\framebox[5cm]{\rule[-25mm]{0cm}{5cm}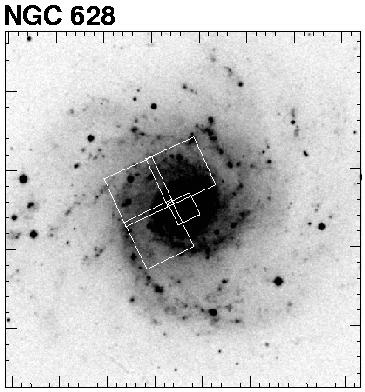 (NGC628)}
\framebox[5cm]{\rule[-25mm]{0cm}{5cm}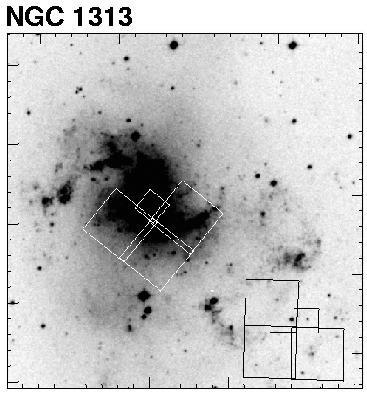 (NGC1313)}
\framebox[5cm]{\rule[-25mm]{0cm}{5cm}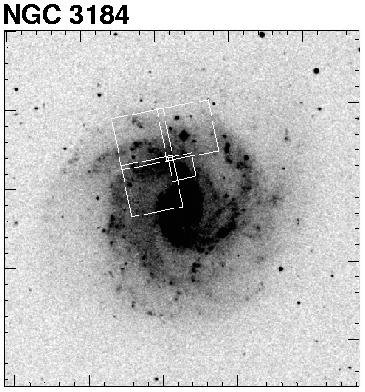 (NGC3184)}\\
\noindent
\framebox[5cm]{\rule[-25mm]{0cm}{5cm}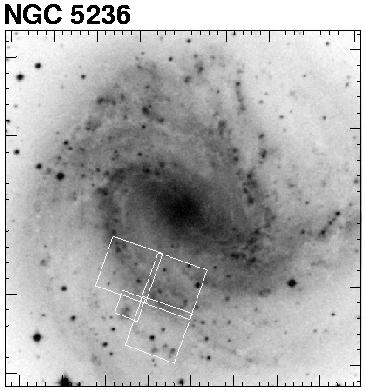 (NGC5236)}
\framebox[5cm]{\rule[-25mm]{0cm}{5cm}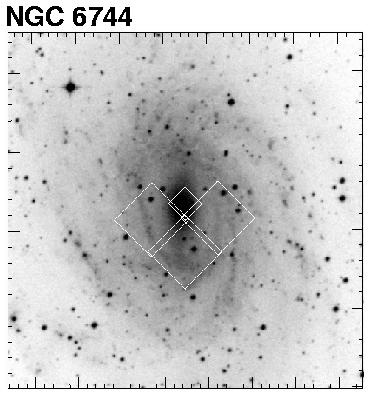 (NGC6744)}
\framebox[5cm]{\rule[-25mm]{0cm}{5cm}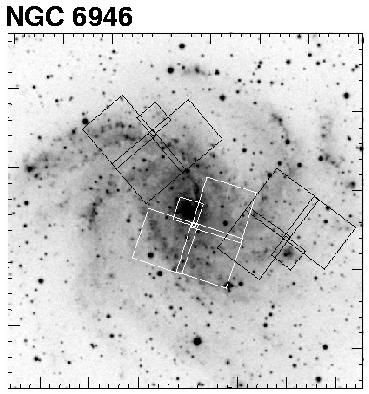 (NGC6946)}
\figcaption{\label{fig:pointings}WFPC2 pointings used in this paper 
 superimposed on images from the Digitized Sky Survey. Each panel covers 
 $9\arcmin\times9\arcmin$.}

\clearpage
\epsfxsize=15cm
\epsfbox{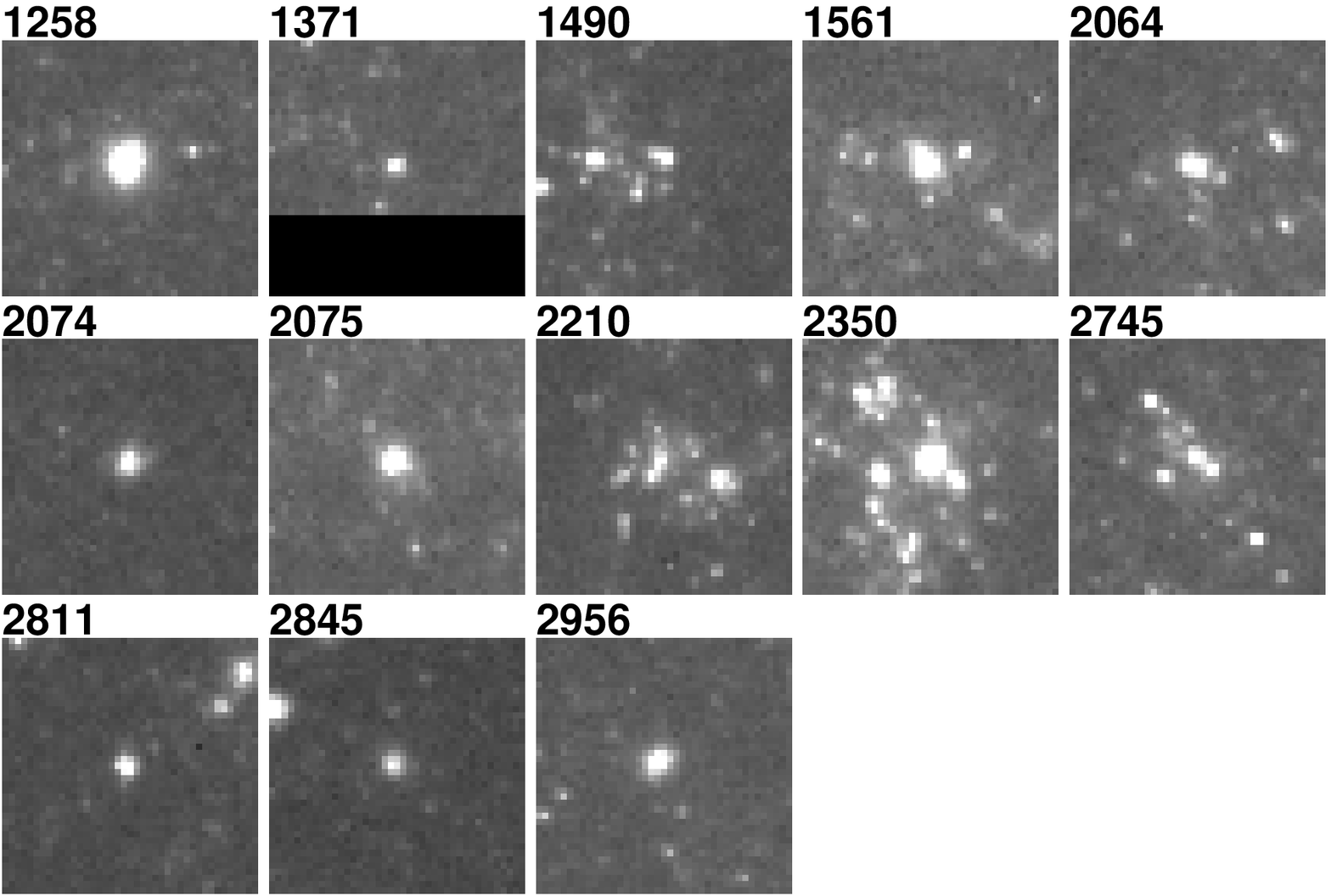}
\figcaption[Larsen.fig2.eps]{\label{fig:cl6946f5}HST images of cluster 
  candidates in the NGC~6946--3 field identified on ground-based images}

\clearpage
\epsfxsize=16cm
\epsfbox{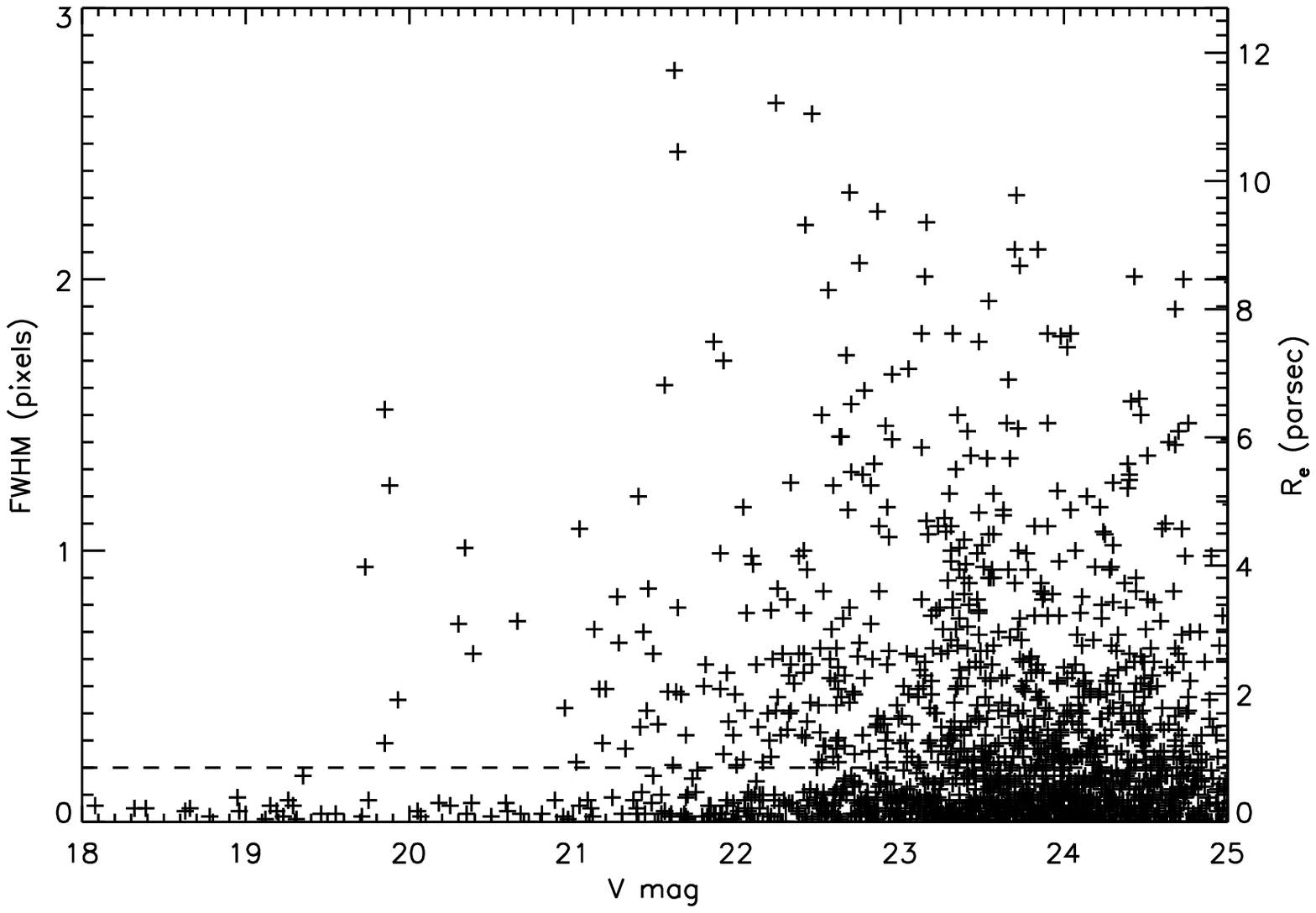}
\figcaption[Larsen.fig3.ps]{\label{fig:v_sz}Intrinsic FWHM in pixels 
  vs.\ $V$ magnitude for sources in field NGC~6946--3.}

\clearpage
\epsfxsize=16cm
\epsfbox{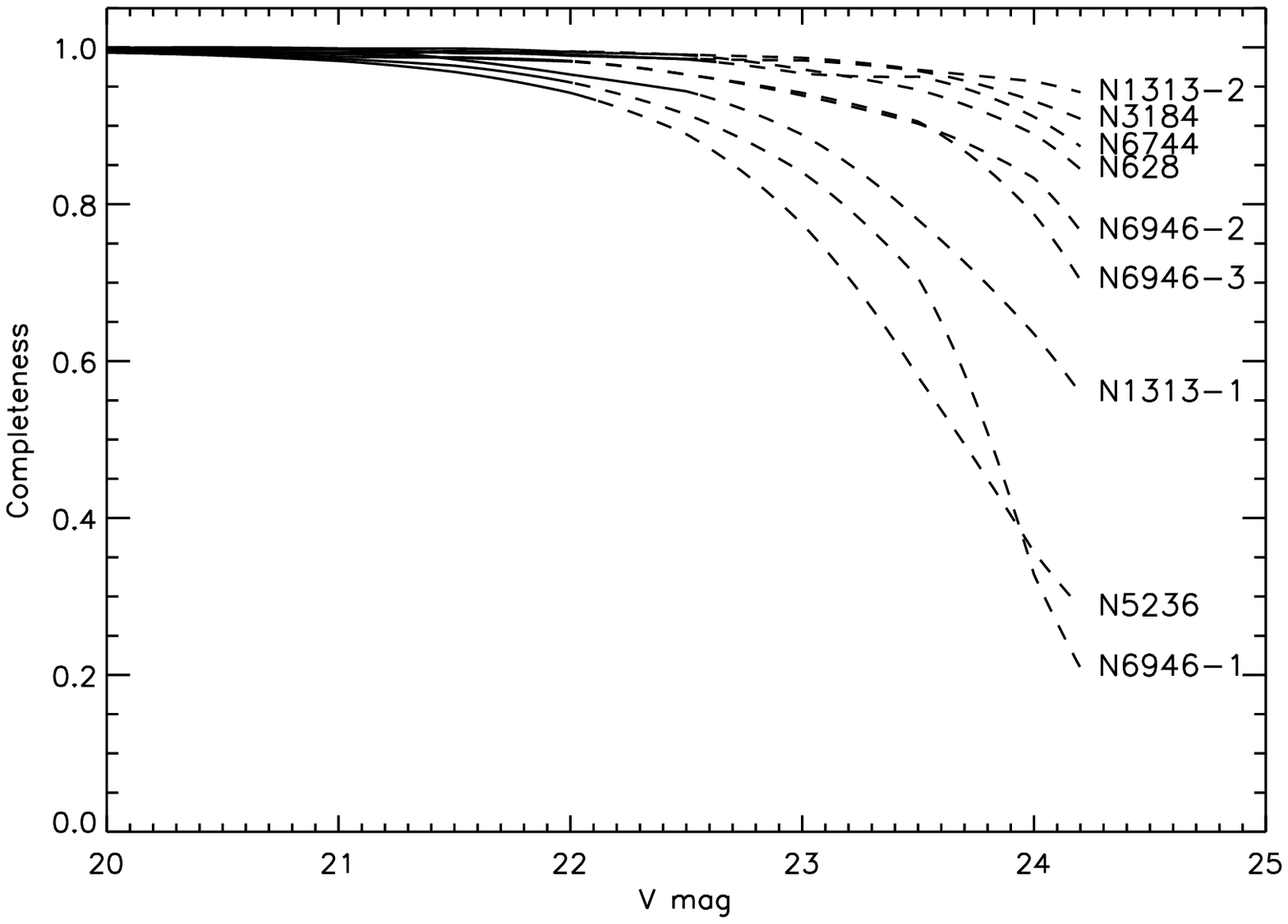}
\figcaption[Larsen.fig4.ps]{\label{fig:cmpl}Completeness functions for
 each of the WFPC2 fields, determined from artificial object experiments. 
 The parts of the curves drawn with solid lines represent the magnitude 
 intervals used for luminosity function fits.}

\clearpage
\epsfxsize=16cm
\epsfbox{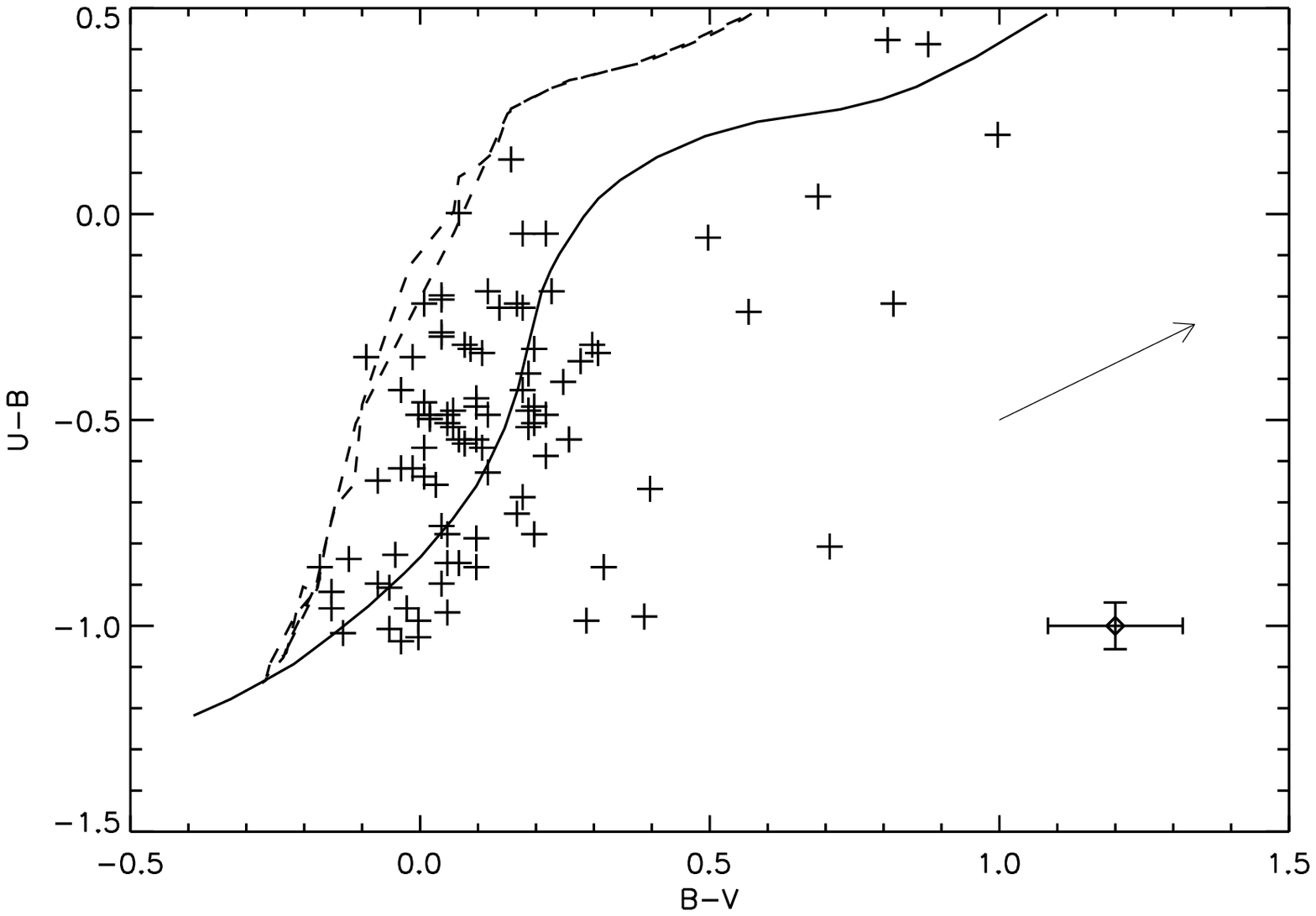}
\figcaption[Larsen.fig5.ps]{\label{fig:bv_ub}(\bv,\ub) two-color diagram
  for cluster candidates in NGC~6946, corrected for Galactic foreground
  reddening. The solid curve is the \citet{gir95} 
  ``S''-sequence, representing the mean color of LMC clusters, while the
  dashed line is a 4 Myr stellar isochrone from \citet{gir02}. A typical
  error bar is shown in the lower right corner and the arrow indicates the 
  reddening vector corresponding to $A_V = 1$ mag.
}

\clearpage
\epsfxsize=16cm
\epsfbox{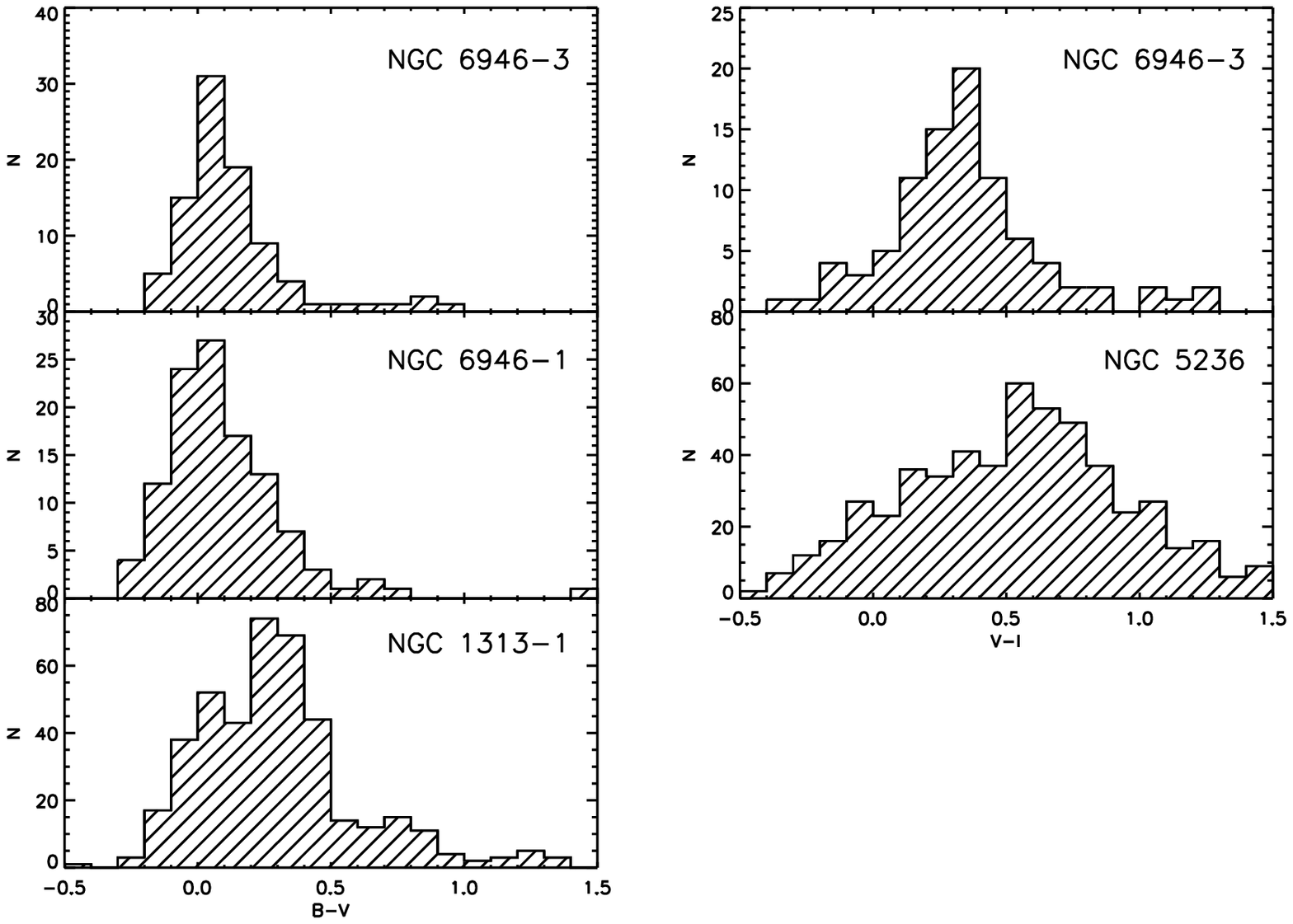}
\figcaption[Larsen.fig6.ps]{\label{fig:colfig}Comparison of \bv\ (left) 
and \vi\ (right) color distributions for cluster candidates in 
fields with color information.}

\clearpage
\epsfxsize=16cm
\epsfbox{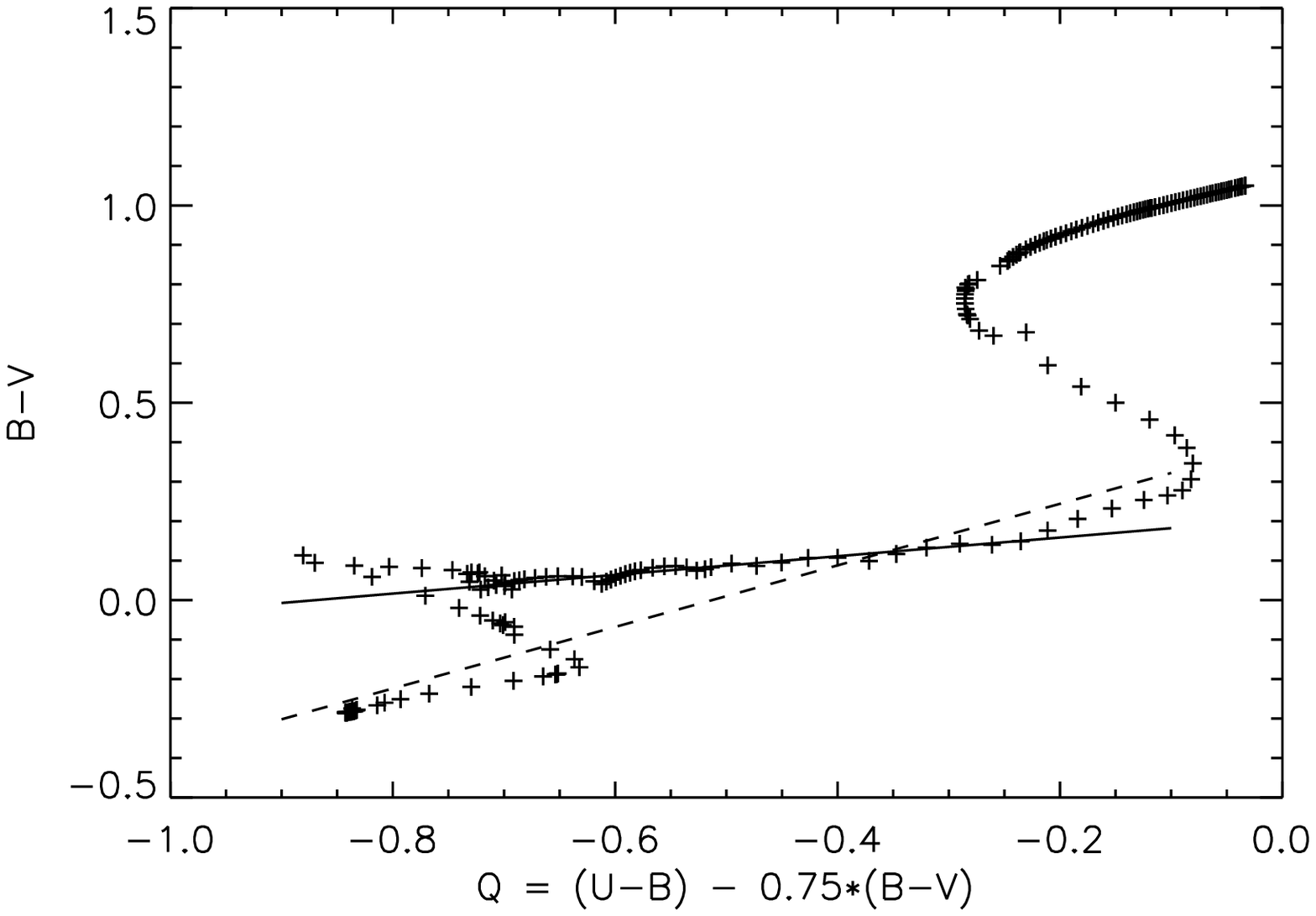}
\figcaption[Larsen.fig7.ps]{\label{fig:q_bv}Reddening-free $Q$ parameter
vs.\ \bv\ according to BC2001 population synthesis models. The dashed line
is the \citet{vh68} relation, while the solid line is a fit to the model
data in the range $-0.6 < Q < -0.2$ and $\bv < 0.3$.}

\clearpage
\epsfxsize=8cm
\epsfbox{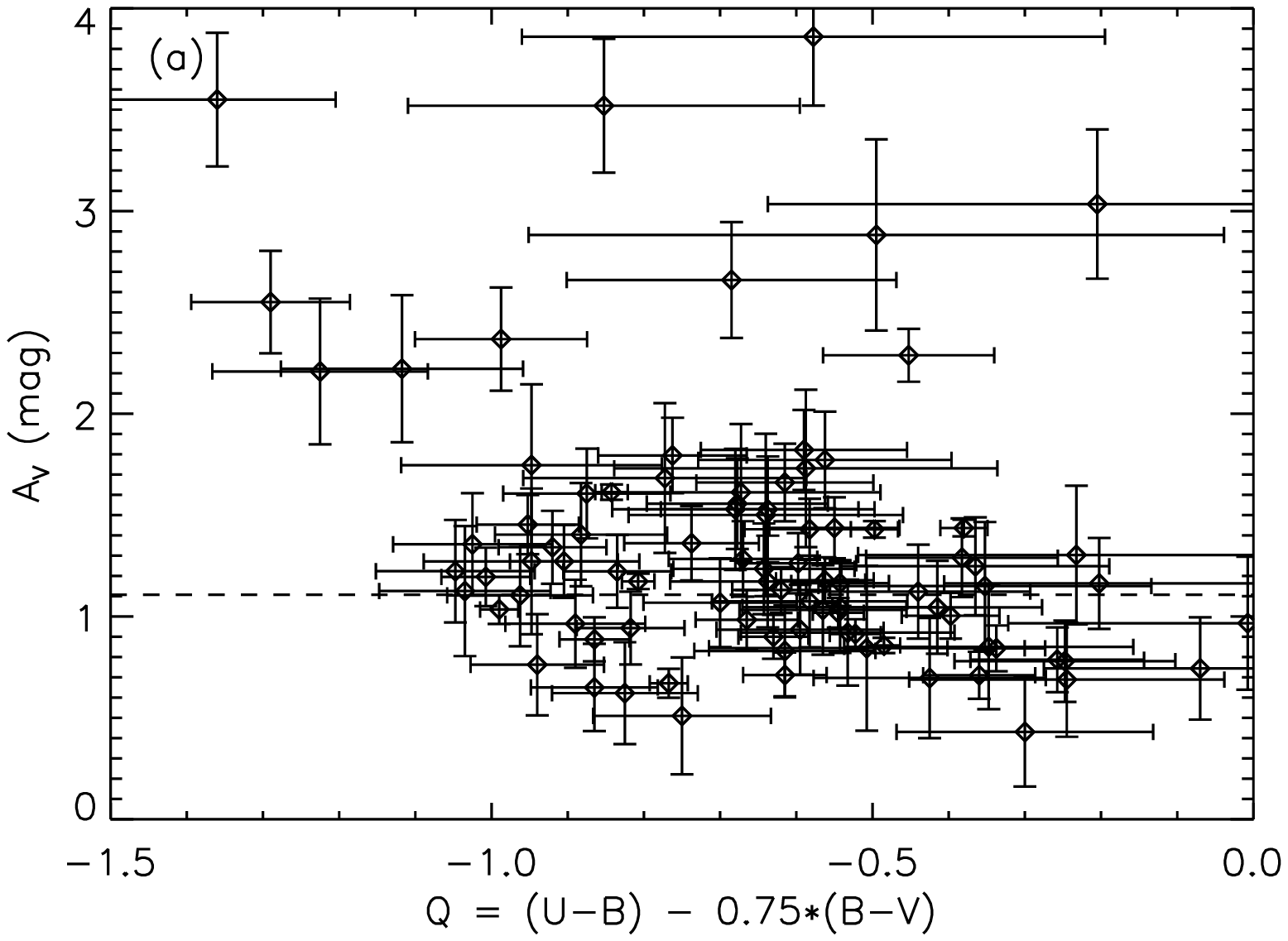}
\epsfxsize=8cm
\epsfbox{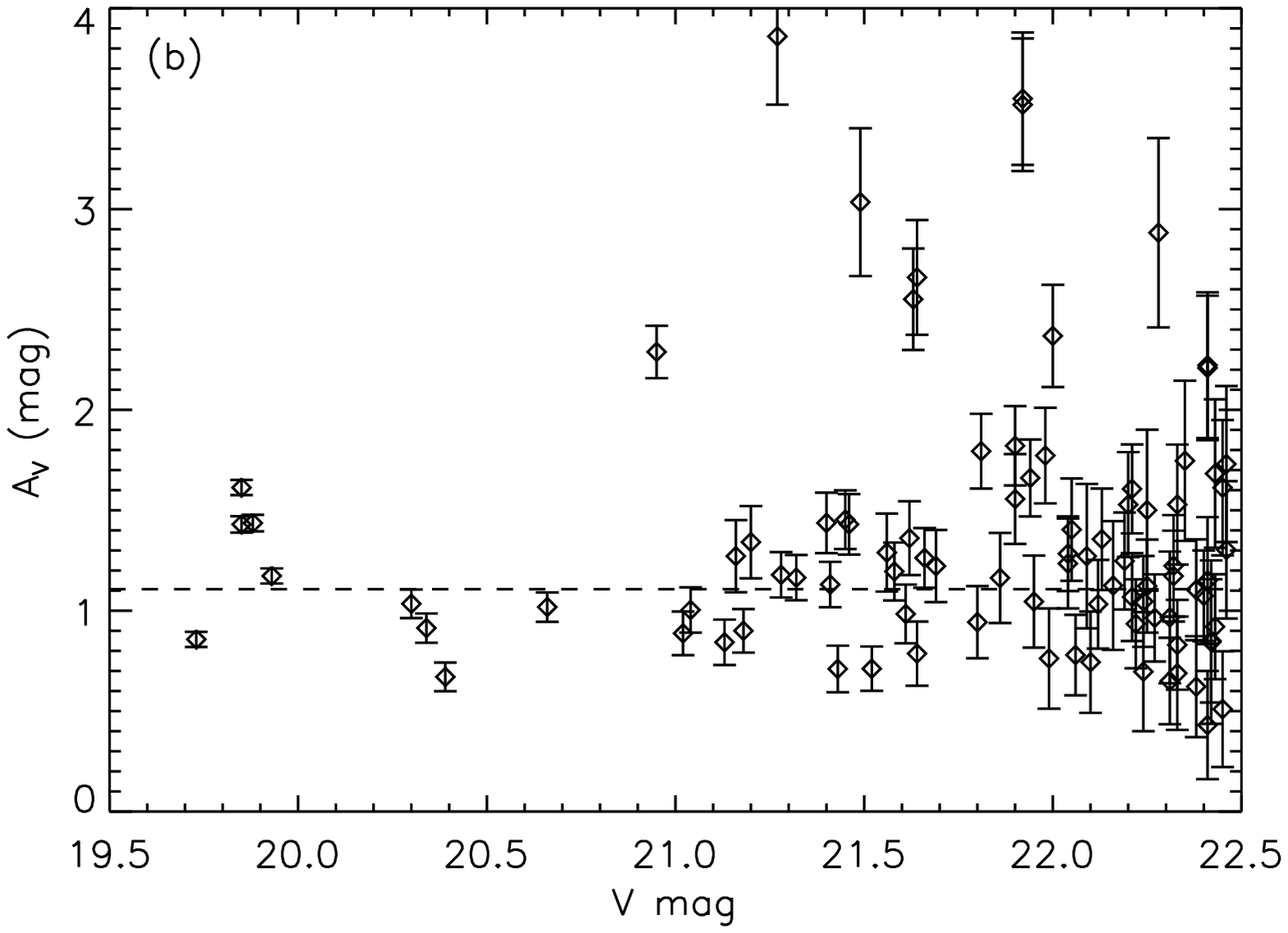}
\figcaption[Larsen.fig8a.ps,Larsen.fig8b]
  {\label{fig:ab}Total extinction (in $V$ mag) 
towards individual clusters in NGC~6946 listed in Table~\ref{tab:cl6946f5}, 
determined from the Q-method. The $A_V$ values are shown
as a function of $Q$-parameter (a) and $V$ magnitude (b). The dashed line 
indicates the Galactic foreground reddening according to \citet{sch98}.}

\clearpage
\begin{minipage}{16.5cm}
\begin{minipage}{8cm}
\epsfxsize=8cm
\epsfbox{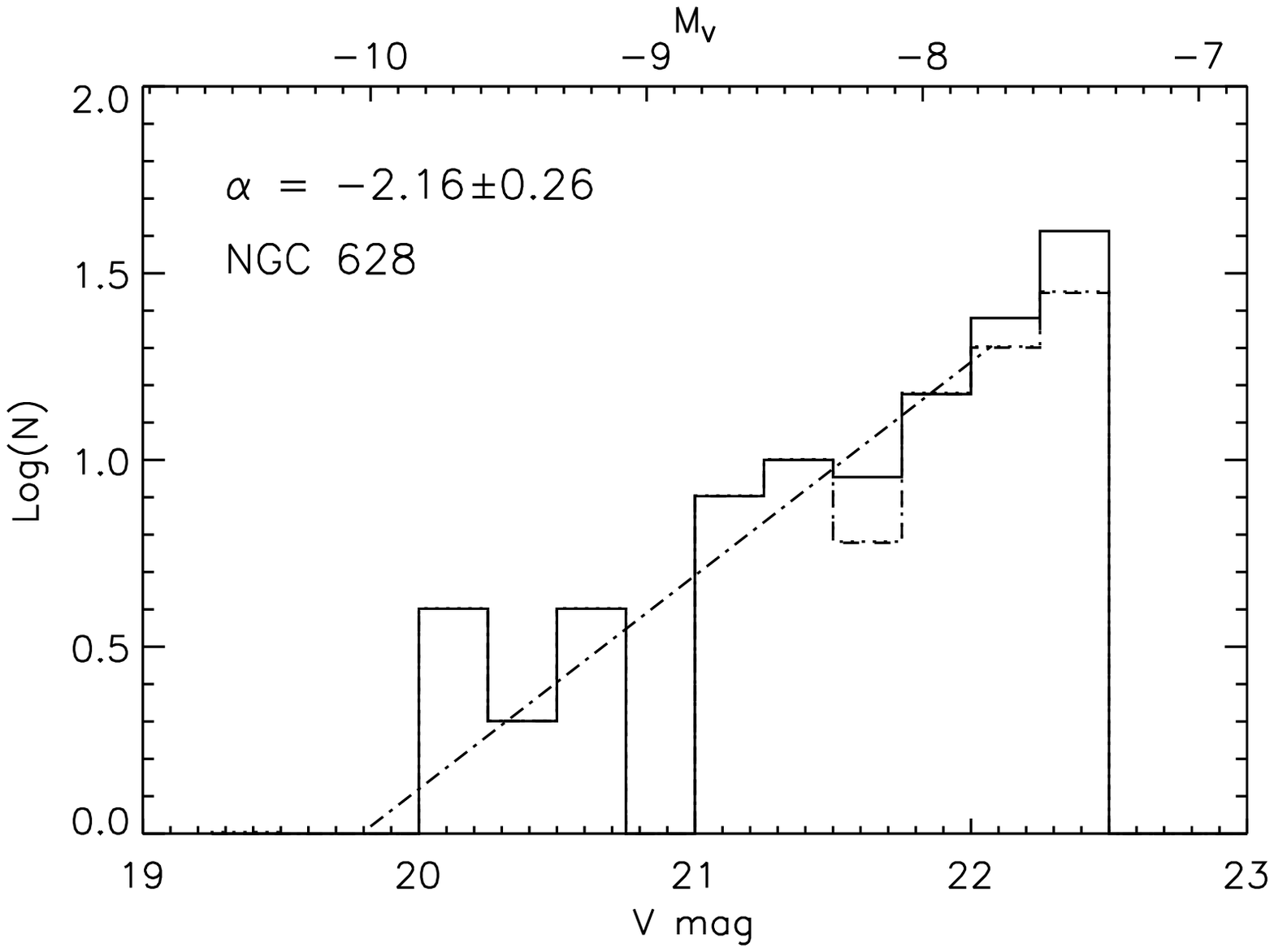}
\end{minipage}
\begin{minipage}{8cm}
\epsfxsize=8cm
\epsfbox{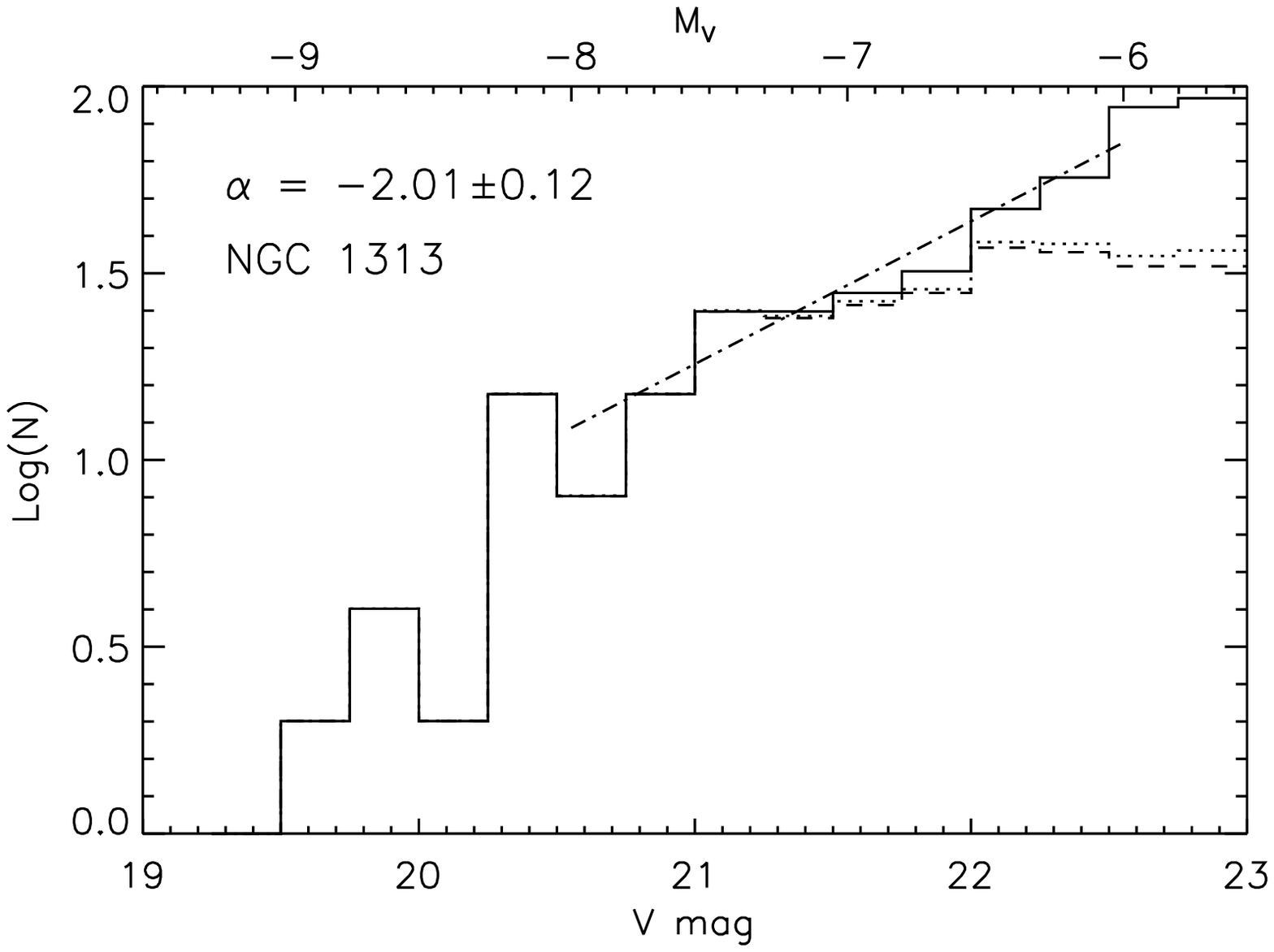}
\end{minipage}
\\

\begin{minipage}{8cm}
\epsfxsize=8cm
\epsfbox{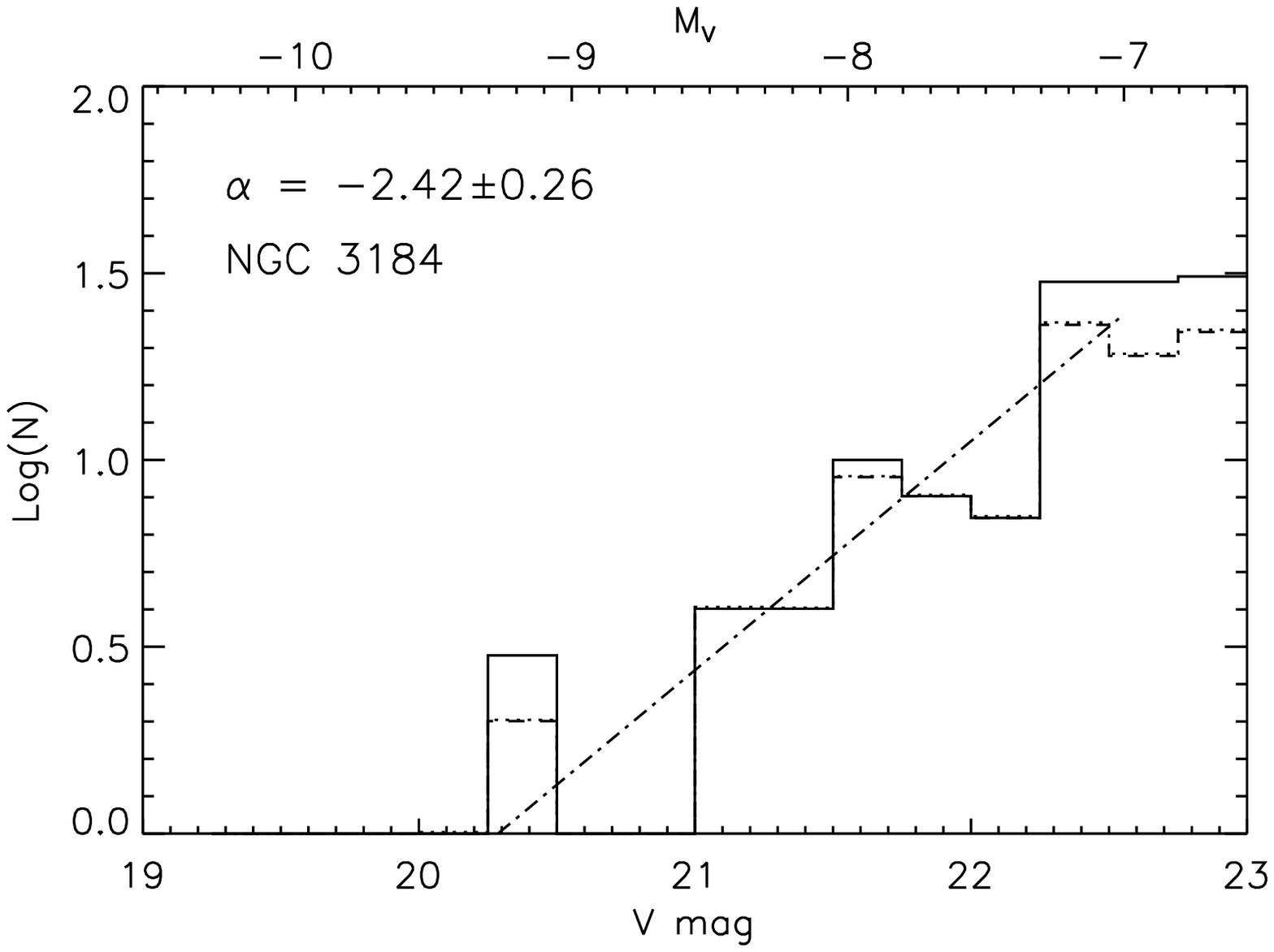}
\end{minipage}
\begin{minipage}{8cm}
\epsfxsize=8cm
\epsfbox{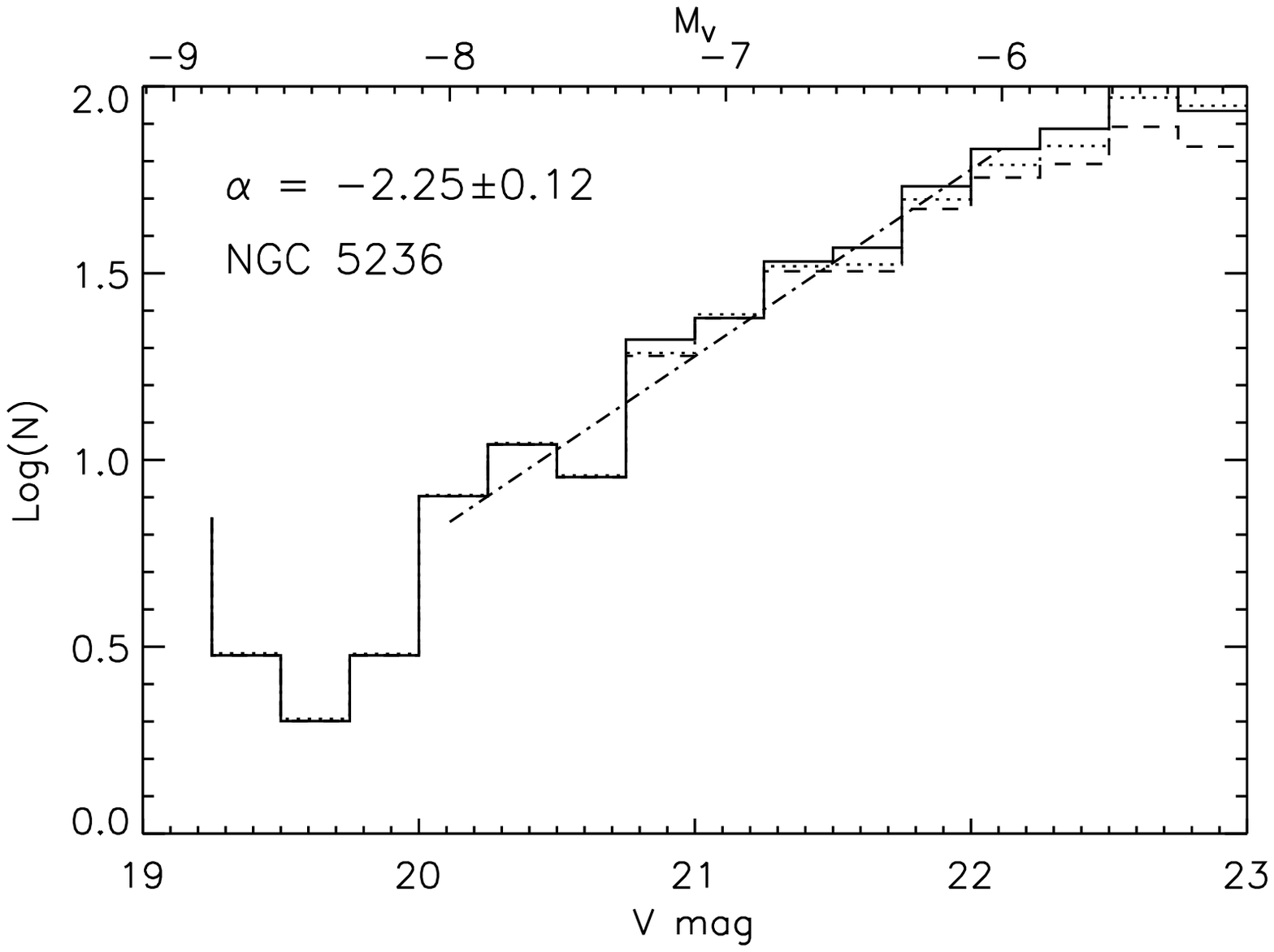}
\end{minipage}
\end{minipage}
\figcaption[Larsen.fig9a.ps,Larsen.fig9b.ps,Larsen.fig9c.ps,Larsen.fig9d.ps]
{\label{fig:lf1}$V$-band luminosity functions for cluster candidates
  in the galaxies. Dotted and dashed lines show the luminosity functions 
  after removal of potential contaminants, with and without correction for
  incompleteness, while the solid lines are the uncorrected LFs.  The 
  dotted-dashed lines represent power-law fits of the form 
  $dN(L)/dL \propto L^{\alpha}$ to the uncorrected LFs, where the $\alpha$ 
  values are shown in each panel.
}

\clearpage
\begin{minipage}{16.5cm}
\begin{minipage}{8cm}
\epsfxsize=8cm
\epsfbox{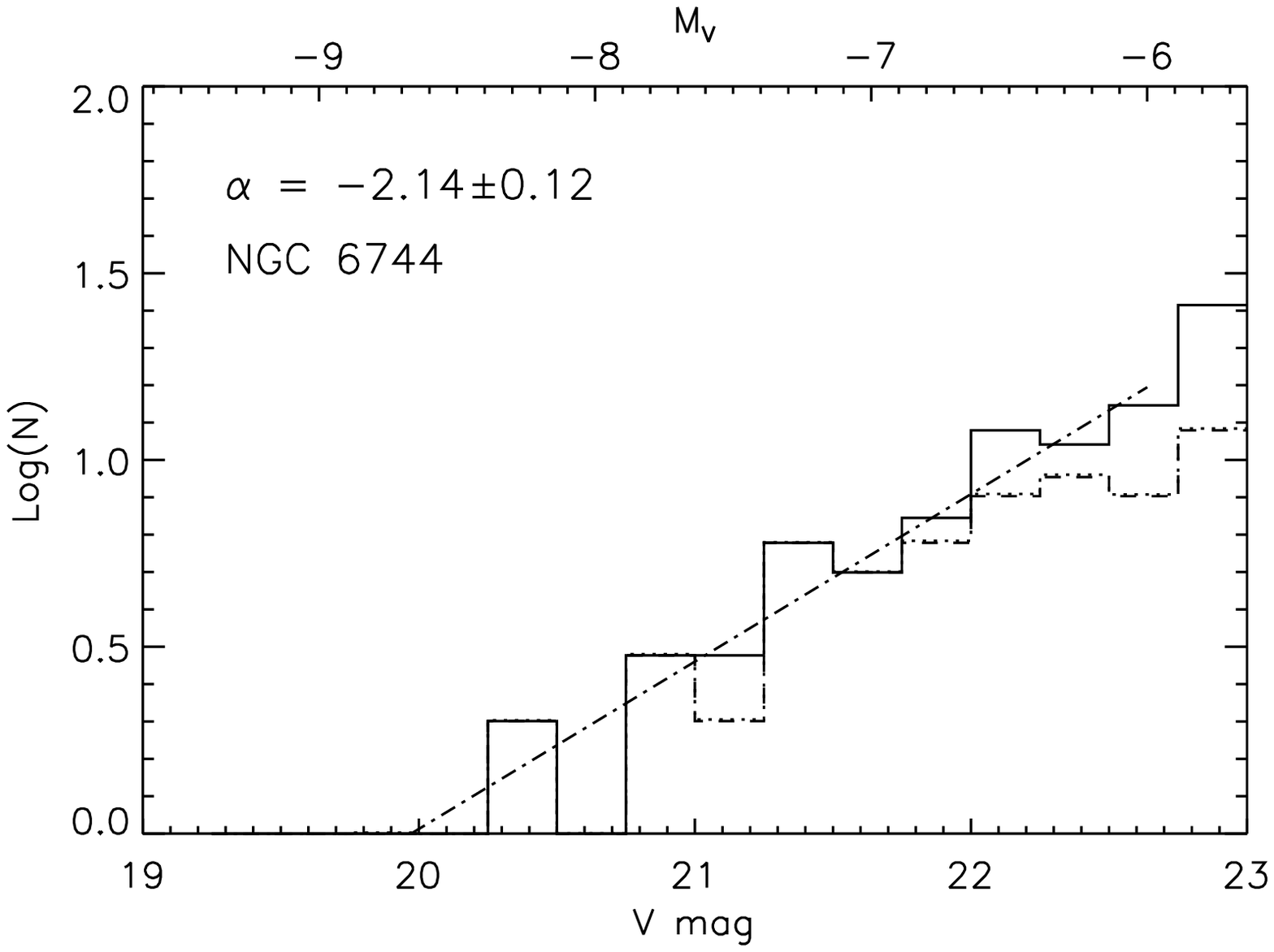}
\end{minipage}
\begin{minipage}{8cm}
\epsfxsize=8cm
\epsfbox{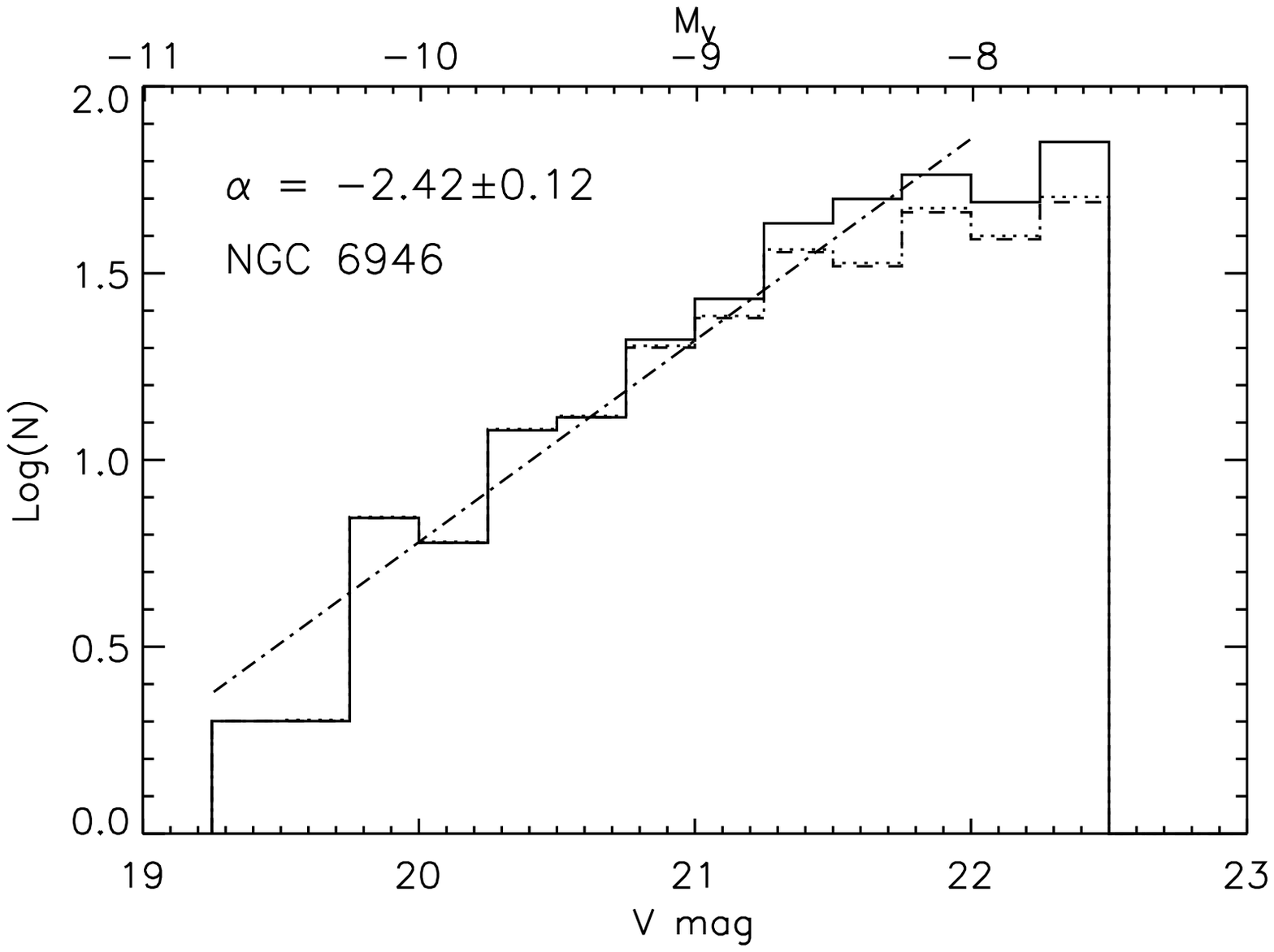}
\end{minipage}
\\
\begin{minipage}{8cm}
\epsfxsize=8cm
\epsfbox{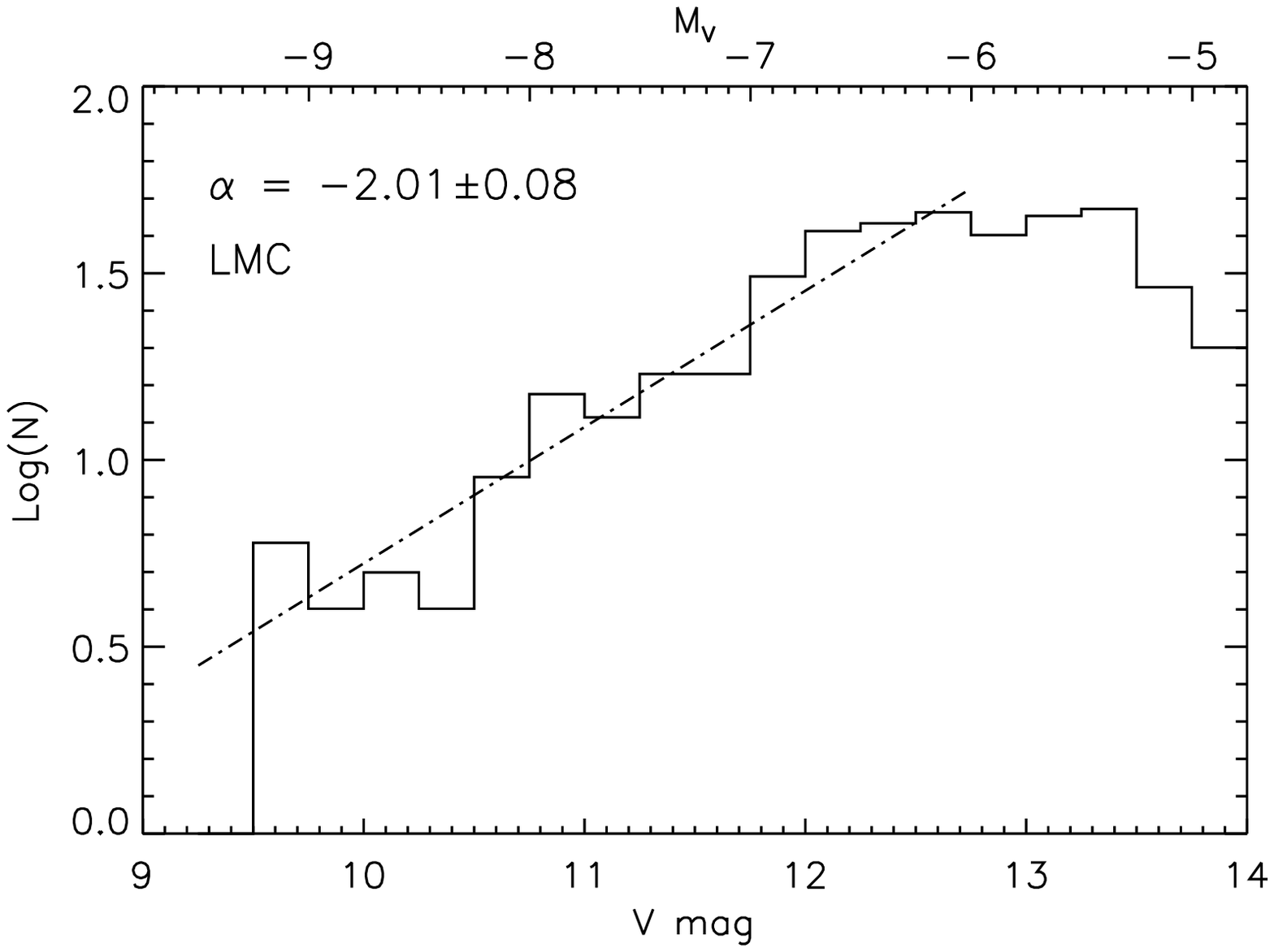}
\end{minipage}
\end{minipage}
\figcaption[Larsen.fig10a.ps,Larsen.fig10b.ps,Larsen.fig10c.ps]
  {\label{fig:lf2}See caption to Fig.~\ref{fig:lf1}}

\clearpage
\epsfxsize=16cm
\epsfbox{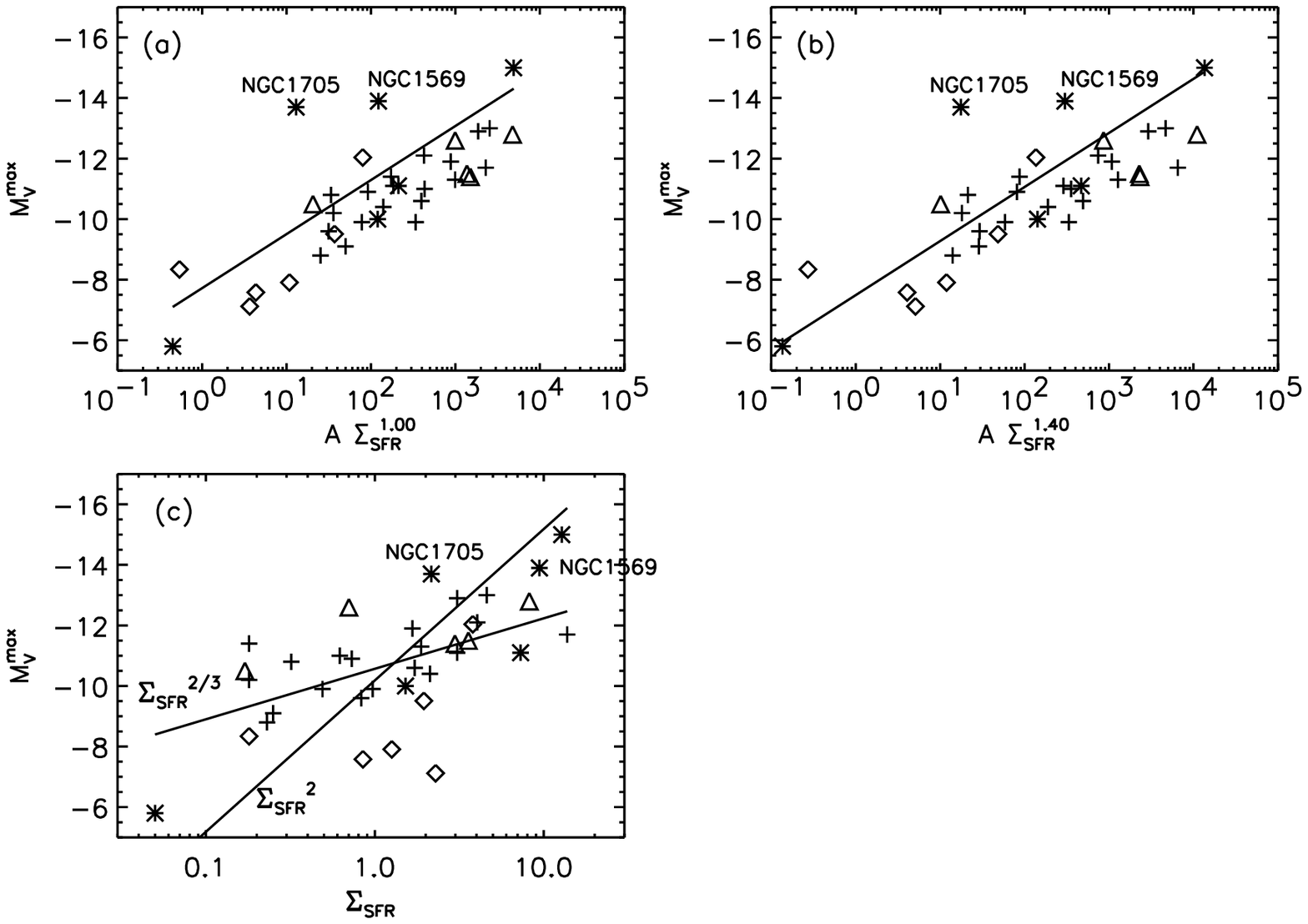}
\figcaption[Larsen.fig11.ps]
{\label{fig:sfr_vm}Comparison of maximum observed 
cluster magnitudes with predictions.  Panels (a) and (b): maximum magnitudes 
versus $A \,\ssfr$ and $A\,\ssfr^{1.4}$. Lines represent predictions 
based on sample statistics for cluster luminosity function with 
$\alpha=-2.4$.  Panel (c): maximum magnitude versus $\ssfr$, with lines 
representing maximum cluster magnitudes for constant cluster density 
($\eta=-2$) or radius ($\eta=2/3$), according to BHE02. Normalizations of
the theoretical relations in panels (a) and (b) come 
from the data in Table~\ref{tab:lffit} but are arbitrary in panel (c).
Symbols : ($+$) -- LR2000, ($*$) -- LR2000, literature data, 
($\diamondsuit$) -- BHE02, ($\triangle$) -- Lick data.
}

\clearpage
\epsfxsize=16cm
\epsfbox{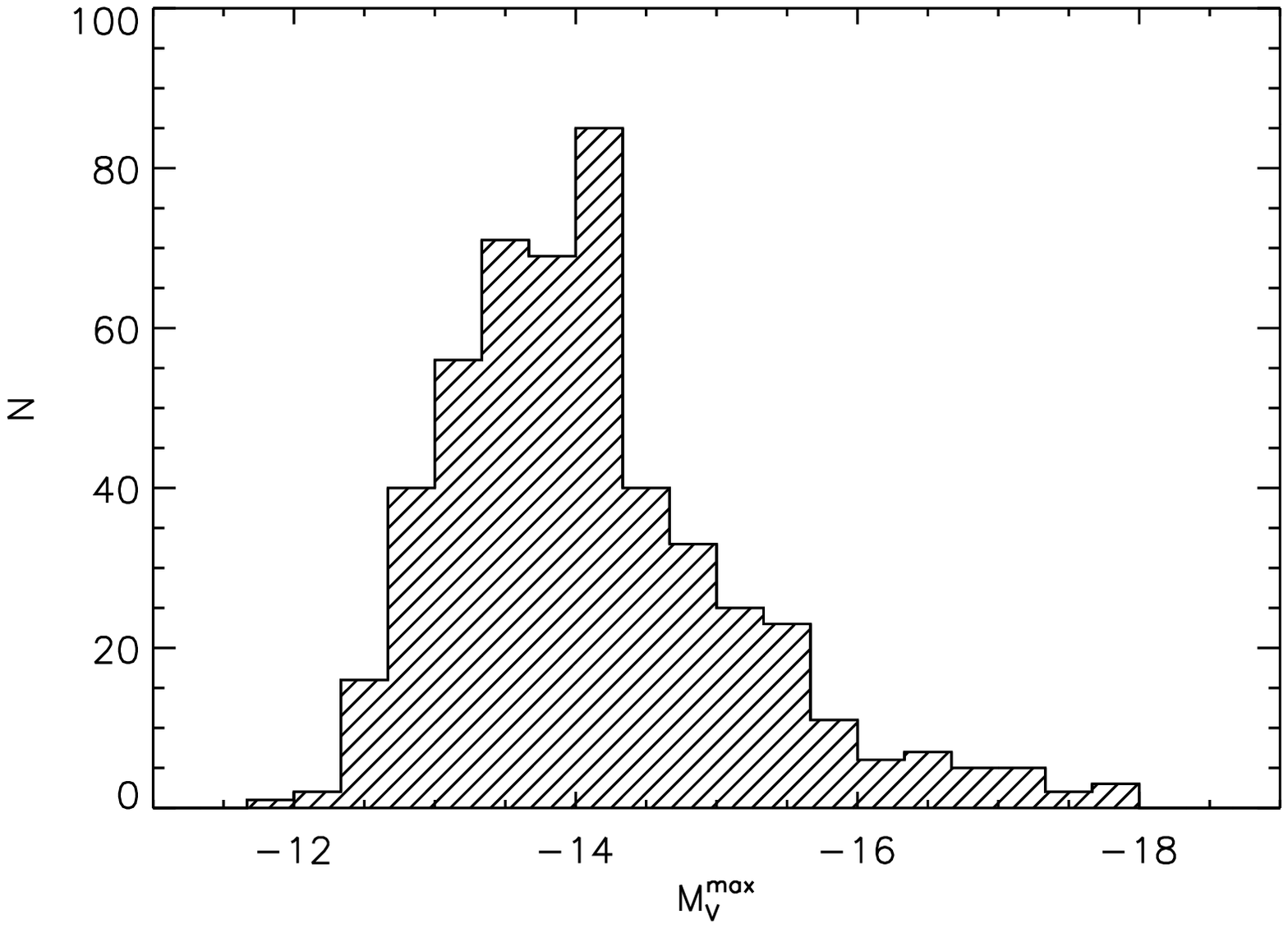}
\figcaption[Larsen.fig12.ps]{\label{fig:lmaxmc}Histogram of brightest absolute 
cluster magnitude for 500 Monte Carlo simulations, assuming random sampling of 
a power-law luminosity function with $\alpha=-2.4$. The cluster population 
was normalized to 2000 clusters per magnitude bin at $M_V=-8$.
}

\clearpage
\begin{minipage}{14cm}
\epsfxsize=14cm
\epsfbox{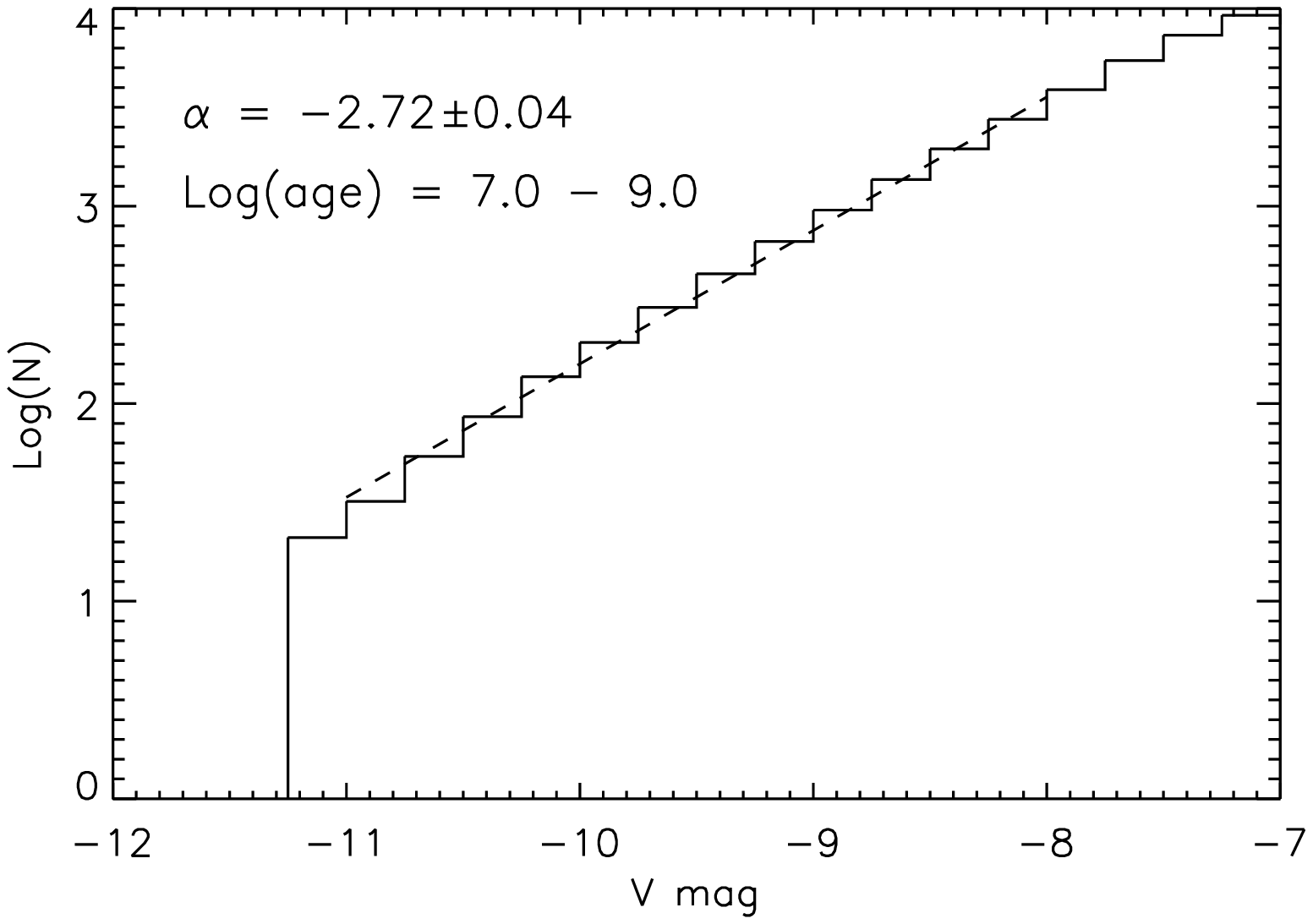}
\end{minipage}
\figcaption[Larsen.fig13.ps]{\label{fig:lf_synt}Simulated luminosity function 
for clusters with ages uniformly distributed over the range $10^7 - 10^9$ 
years.  For the mass function, a power-law with slope $\alpha=-2.0$ in the 
interval $10^3 < M < 10^5 \msun$ was assumed.  The dashed line represents a 
power-law fit to the luminosity function in the range $-11 < M_V < -8$. 
}

\clearpage
\epsfxsize=16cm
\epsfbox{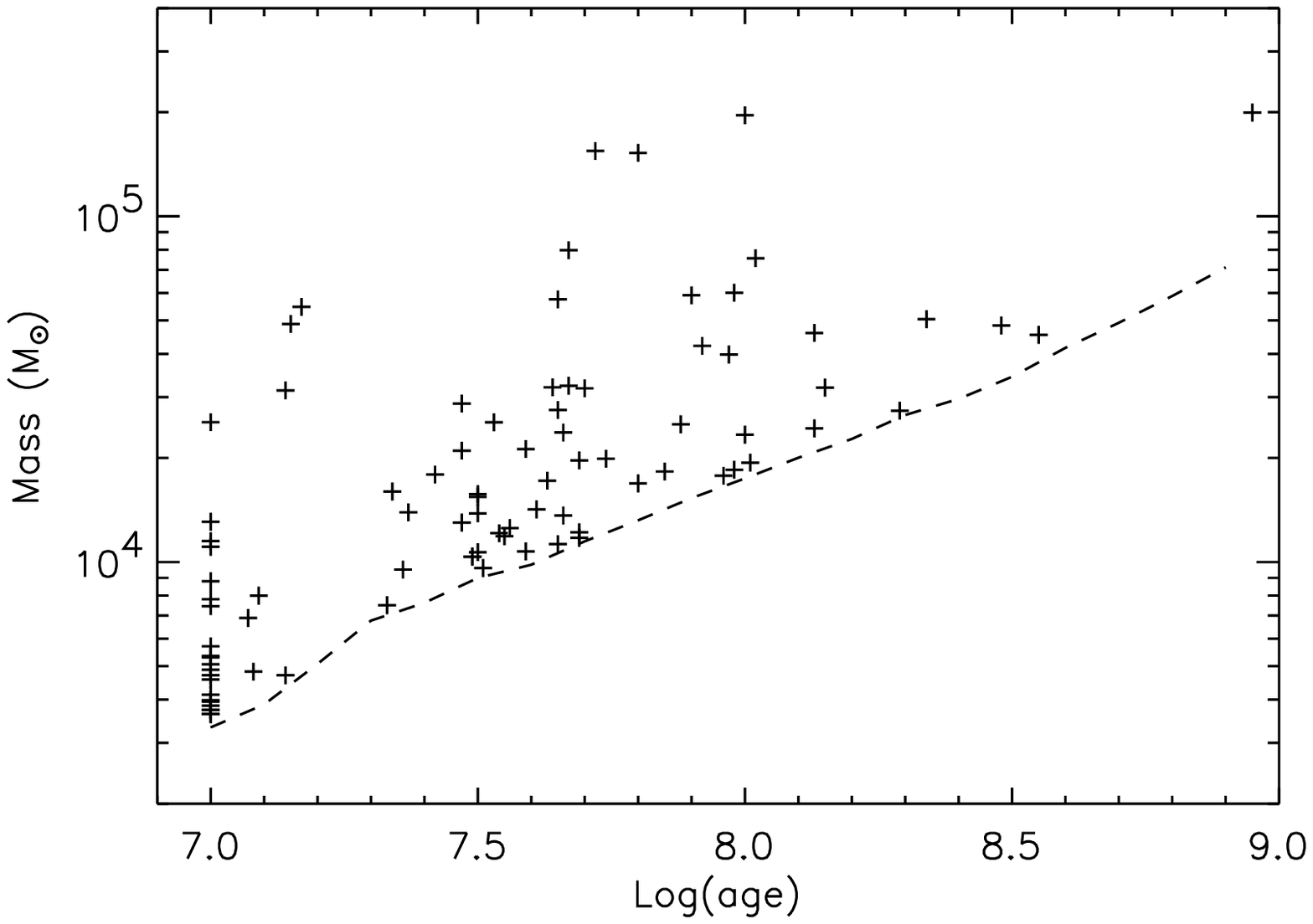}
\figcaption[Larsen.fig14.ps]{\label{fig:age_m}Mass vs.\ age for clusters in 
NGC~6946 with $UBV$ data. The dashed curve indicates the observational 
cut-off as defined by the magnitude limit and age-dependent M/L ratio.}

\clearpage
\begin{deluxetable}{lcccl}
\tablecaption{\label{tab:datasets}The datasets used in this paper.
$A_B$ values are from \citet{sch98}, retrieved from the 
\emph{NASA/IPAC Extragalactic Database}. References for the distance moduli 
$(m-M)_0$ are given in Section~\ref{sec:data}.
}
\tablehead{Field & $A_B$ & $(m-M)_0$ & Program ID &  Exposures}
\startdata
NGC 628     & 0.30 & 29.6 & 8597 & F606W ($160+400$ s) \\
NGC 1313--1 & 0.47 & 28.2 & 9042 & F450W ($2\times230$ s), F606W ($2\times230$ s) \\
NGC 1313--2 &  -   &  -   & 6713 & F606W ($2\times300$ s) \\
NGC 3184    & 0.07 & 29.5 & 8602 & F555W ($2\times350$ s) \\
NGC 5236    & 0.28 & 27.9 & 5971 & F606W ($1100+1200$ s), F814W ($2\times1000$ s) \\
NGC 6744    & 0.19 & 28.5 & 8597 & F606W ($160+400$ s) \\
NGC 6946--1 & 1.48 & 28.9 & 6118 & F439W ($2\times400$ s), F555W (400 s) \\
NGC 6946--2 &  -   &  -   & 8597 & F606W ($160+400$ s) \\
NGC 6946--3 &  -   &  -   & 8715 & F336W ($2\times1500$ s), F439W ($2\times1100$ s) \\
            &      &      &    & F555W ($2\times300$ s), F814W ($2\times700$ s) 
\enddata   
\end{deluxetable}

\begin{deluxetable}{lrrrrrrr}
\tablecaption{\label{tab:apc}Aperture corrections for WF images relative 
  to $r=5$ pixels, determined from measurements on synthetic images.}
\tablehead{$r$  & F336W & F439W & F555W & F814W &  F336W &  F439W &  F555W \\
                &     &     &     &     & $-$F439W  & $-$F555W & $-$F814W }
\startdata
 & \multicolumn{7}{l}{Point sources} \\
%
  3.0 & 0.061   &  0.053   & 0.055   & 0.070   &  0.008  & $-$0.002 & $-$0.015\\
 20.0 & $-0.083$& $-$0.078 &$-$0.070 &$-$0.076 &$-$0.005 & $-$0.008 &  0.006 \\
  & \multicolumn{7}{l}{King $c=30$, FWHM = 0.5 pixels} \\
 3.0  & 0.152   & 0.149    & 0.151   & 0.165   &  0.003  & $-$0.002 &$-$0.014 \\
20.0  & $-$0.104& $-$0.097 & $-$0.090&$-$0.098 &$-$0.007 & $-$0.007 &  0.008 \\
  & \multicolumn{7}{l}{King $c=30$, FWHM = 1.0 pixels} \\
 3.0  & 0.271   & 0.268    &0.270    &0.283    &  0.003  & $-$0.002 &$-$0.013 \\
20.0  &$-$0.226 &$-$0.219  &$-$0.213 &$-$0.222 &$-$0.007 & $-$0.006 &  0.009 \\
  & \multicolumn{7}{l}{MOFFAT15, FWHM = 0.5 pixels} \\
 3.0  & 0.123   & 0.119    &0.121    &0.135    & 0.004   & $-$0.002 &$-$0.014 \\
20.0  &$-$0.137 &$-$0.130  &$-$0.123 &$-$0.130 &$-$0.007 & $-$0.007 &  0.007 \\
  & \multicolumn{7}{l}{MOFFAT15, FWHM = 1.0 pixels} \\
 3.0  & 0.191   & 0.187    & 0.191   &0.203    & 0.004   & $-$0.004 &$-$0.012\\
20.0  &$-$0.194 &$-$0.187  &$-$0.180 &$-$0.189 &$-$0.007 & $-$0.007 &  0.009\\
\enddata
\end{deluxetable}

\begin{deluxetable}{lrrrrrrrrrrr}
\rotate
\tabletypesize{\footnotesize}
\tablewidth{0pt}
\tablecaption{\label{tab:f5}Comparison of ground-based and HST data
  for clusters in NGC~6946. No corrections for reddening have been applied.
  Effective radii (\reff) are in pc. }
\tablehead{ID (L99) & \multicolumn{5}{c|}{Ground} & \multicolumn{6}{c}{HST} \\
        &   $V$ & \ub & \bv & \vi & \multicolumn{1}{c|}{\reff }
	  & $V$ & $V_{15}$ & \ub & \bv & \vi & \reff}
\startdata
1258   & 19.50 & $ 0.02$ &  0.61 &  1.00 &  5.9 & $19.77\pm0.01$ & $19.46\pm0.01$ & $ 0.05\pm0.03$ & $ 0.58\pm0.01$ & $ 1.05\pm0.01$ &  4.9 \\
1371   & 21.07 & $ 0.11$ &  0.66 &  0.91 &  6.6 & $21.61\pm0.03$ &       -        & $ 0.14\pm0.21$ & $ 0.63\pm0.06$ & $ 0.90\pm0.04$ &  0.9 \\
1561   & 19.49 & $-0.49$ &  0.38 &  0.75 &  7.2 & $19.84\pm0.01$ & $19.49\pm0.02$ & $-0.51\pm0.02$ & $ 0.42\pm0.01$ & $ 0.71\pm0.01$ &  1.5 \\
2064   & 19.94 & $-0.63$ &  0.26 &  0.54 &  6.5 & $20.27\pm0.01$ & $19.88\pm0.02$ & $-0.61\pm0.02$ & $ 0.26\pm0.02$ & $ 0.52\pm0.02$ &  2.0 \\
2074   & 21.07 & $-0.08$ &  0.45 &  0.89 &  6.9 & $21.27\pm0.02$ & $20.98\pm0.05$ & $-0.20\pm0.09$ & $ 0.60\pm0.04$ & $ 0.97\pm0.02$ &  4.0 \\
2075   & 20.48 & $-0.02$ &  0.71 &  1.26 &  9.9 & $20.84\pm0.02$ & $20.52\pm0.04$ & $ 0.20\pm0.14$ & $ 0.84\pm0.04$ & $ 1.42\pm0.02$ &  1.7 \\
2350   & 18.52 & $-0.58$ &  0.35 &  0.63 &  8.9 & $19.17\pm0.01$ & $18.29\pm0.02$ & $-0.61\pm0.03$ & $ 0.29\pm0.02$ & $ 0.63\pm0.03$ &  0.5 \\
2811   & 21.27 & $ 0.00$ &  0.45 &  1.07 & 14.5 & $21.65\pm0.03$ & $21.51\pm0.09$ & $-0.17\pm0.11$ & $ 0.47\pm0.05$ & $ 1.05\pm0.03$ &  1.8 \\
2845   & 21.07 & $ 0.06$ &  0.38 &  0.90 & 12.6 & $21.51\pm0.02$ & $21.37\pm0.07$ & $ 0.02\pm0.12$ & $ 0.43\pm0.04$ & $ 0.87\pm0.03$ &  2.9 \\
2956   & 20.77 & $-0.07$ &  0.39 &  0.92 &  6.4 & $20.95\pm0.01$ & $20.67\pm0.04$ & $-0.05\pm0.07$ & $ 0.43\pm0.03$ & $ 0.91\pm0.02$ &  3.7 \\
\enddata
\end{deluxetable}

\begin{deluxetable}{lrrrrrrrrc}
\tabletypesize{\footnotesize}
\tablewidth{0pt}
\tablecaption{\label{tab:cl6946f5}Cluster candidates in field NGC~6946--3 
selected from HST images. No correction for reddening has been
applied to the photometry in the table.}
\tablehead{chip/ID & x & y & $V$ & \ub & \bv & \vi & FWHM   & log(age) & Mass \\
                   &   &   &     &     &     &     & pixels &  years & $10^3$ \msun}
\startdata
WF2/50 & 395 & 81 & $22.19\pm0.03$ & $0.04\pm0.17$ & $0.54\pm0.06$ & $0.95\pm0.04$ & 0.40 & $8.00\pm0.29$ & $22\pm11$ \\
WF2/63 & 321 & 93 & $22.38\pm0.03$ & $-0.70\pm0.08$ & $0.35\pm0.07$ & $0.25\pm0.06$ & 0.62 & $<7.0$ & $3\pm2$ \\
WF2/113 & 464 & 147 & $22.32\pm0.03$ & $-0.30\pm0.12$ & $0.45\pm0.06$ & $0.86\pm0.04$ & 0.54 & $7.50\pm0.26$ & $10\pm4$ \\
WF2/172 & 200 & 226 & $22.25\pm0.03$ & $-0.08\pm0.14$ & $0.48\pm0.06$ & $1.01\pm0.04$ & 0.46 & $7.85\pm0.25$ & $17\pm7$ \\
WF2/211 & 70 & 317 & $22.24\pm0.03$ & $-0.22\pm0.12$ & $0.43\pm0.06$ & $1.00\pm0.04$ & 0.41 & $7.66\pm0.22$ & $14\pm5$ \\
WF2/226 & 430 & 340 & $19.85\pm0.01$ & $-0.07\pm0.03$ & $0.57\pm0.01$ & $0.98\pm0.01$ & 1.52 & $7.80\pm0.14$ & $153\pm36$ \\
WF2/233 & 622 & 344 & $22.32\pm0.03$ & $-0.77\pm0.09$ & $0.37\pm0.07$ & $0.39\pm0.06$ & 0.40 & $<7.0$ & $3\pm3$ \\
WF2/262 & 254 & 383 & $21.94\pm0.02$ & $-0.15\pm0.11$ & $0.62\pm0.05$ & $0.94\pm0.03$ & 0.55 & $7.63\pm0.21$ & $18\pm6$ \\
WF2/263 & 653 & 383 & $22.21\pm0.03$ & $-0.47\pm0.10$ & $0.54\pm0.06$ & $1.20\pm0.04$ & 0.24 & $7.08\pm0.30$ & $5\pm2$ \\
WF2/278 & 549 & 415 & $22.45\pm0.04$ & $-0.60\pm0.10$ & $0.20\pm0.08$ & $0.76\pm0.06$ & 5.47 & $7.14\pm0.27$ & $4\pm2$ \\
WF2/299 & 221 & 449 & $22.40\pm0.03$ & $-0.26\pm0.12$ & $0.43\pm0.06$ & $0.98\pm0.04$ & 0.32 & $7.59\pm0.23$ & $11\pm4$ \\
WF2/338 & 279 & 512 & $20.34\pm0.01$ & $-0.23\pm0.03$ & $0.39\pm0.02$ & $0.81\pm0.01$ & 1.01 & $7.67\pm0.14$ & $82\pm19$ \\
WF2/357 & 700 & 530 & $21.86\pm0.02$ & $0.21\pm0.20$ & $0.55\pm0.05$ & $0.84\pm0.03$ & 1.77 & $8.34\pm0.46$ & $49\pm39$ \\
WF2/379 & 359 & 544 & $22.21\pm0.03$ & $-0.40\pm0.09$ & $0.40\pm0.06$ & $1.15\pm0.04$ & 0.78 & $7.36\pm0.21$ & $9\pm3$ \\
WF2/398 & 144 & 564 & $22.06\pm0.03$ & $0.06\pm0.14$ & $0.41\pm0.05$ & $0.96\pm0.03$ & 0.77 & $8.15\pm0.29$ & $31\pm15$ \\
WF2/469 & 625 & 645 & $21.63\pm0.02$ & $-0.72\pm0.09$ & $0.76\pm0.07$ & $0.72\pm0.04$ & 0.48 & $<7.0$ & $7\pm1$ \\
WF2/470 & 626 & 647 & $21.58\pm0.02$ & $-0.73\pm0.05$ & $0.37\pm0.04$ & $0.65\pm0.03$ & 0.48 & $<7.0$ & $7\pm3$ \\
WF2/503 & 77 & 673 & $19.73\pm0.01$ & $-0.20\pm0.02$ & $0.38\pm0.01$ & $0.81\pm0.01$ & 0.94 & $7.72\pm0.14$ & $154\pm36$ \\
WF2/544 & 742 & 705 & $22.31\pm0.04$ & $-0.70\pm0.07$ & $0.22\pm0.06$ & $0.47\pm0.05$ & 0.34 & $<7.0$ & $3\pm1$ \\
WF2/554 & 718 & 713 & $22.13\pm0.03$ & $-0.71\pm0.09$ & $0.42\pm0.07$ & $1.49\pm0.04$ & 0.35 & $<7.0$ & $4\pm3$ \\
WF2/589 & 431 & 744 & $22.04\pm0.03$ & $-0.31\pm0.09$ & $0.48\pm0.05$ & $1.21\pm0.03$ & 0.23 & $7.47\pm0.19$ & $13\pm4$ \\
WF3/47 & 130 & 108 & $22.46\pm0.04$ & $-0.10\pm0.24$ & $0.65\pm0.10$ & $0.92\pm0.05$ & 2.61 & $7.69\pm0.38$ & $12\pm7$ \\
WF3/82 & 130 & 166 & $21.40\pm0.02$ & $-0.13\pm0.08$ & $0.56\pm0.04$ & $0.99\pm0.02$ & 1.20 & $7.70\pm0.20$ & $32\pm11$ \\
WF3/104 & 604 & 185 & $22.27\pm0.03$ & $-0.65\pm0.08$ & $0.32\pm0.06$ & $0.75\pm0.05$ & 0.32 & $<7.0$ & $4\pm2$ \\
WF3/123 & 614 & 206 & $21.64\pm0.02$ & $0.05\pm0.11$ & $0.41\pm0.04$ & $0.89\pm0.03$ & 0.79 & $8.13\pm0.28$ & $44\pm21$ \\
WF3/128 & 546 & 216 & $21.95\pm0.03$ & $-0.07\pm0.13$ & $0.46\pm0.06$ & $0.98\pm0.04$ & 0.37 & $7.88\pm0.24$ & $24\pm9$ \\
WF3/130 & 547 & 221 & $21.28\pm0.02$ & $-0.21\pm0.06$ & $0.47\pm0.03$ & $0.83\pm0.02$ & 0.66 & $7.64\pm0.16$ & $33\pm9$ \\
WF3/137 & 512 & 227 & $21.92\pm0.03$ & $0.04\pm0.25$ & $1.19\pm0.08$ & $1.62\pm0.03$ & 0.25 & $7.50\pm0.48$ & $15\pm12$ \\
WF3/151 & 570 & 244 & $22.22\pm0.03$ & $-0.31\pm0.10$ & $0.38\pm0.06$ & $0.80\pm0.04$ & 0.60 & $7.54\pm0.23$ & $12\pm4$ \\
WF3/190 & 522 & 286 & $21.43\pm0.02$ & $-0.09\pm0.07$ & $0.36\pm0.03$ & $0.89\pm0.02$ & 0.70 & $7.92\pm0.17$ & $41\pm11$ \\
WF3/239 & 515 & 332 & $21.46\pm0.02$ & $-0.17\pm0.08$ & $0.55\pm0.04$ & $1.04\pm0.03$ & 0.86 & $7.65\pm0.18$ & $28\pm8$ \\
WF3/245 & 512 & 338 & $22.33\pm0.04$ & $-0.26\pm0.15$ & $0.56\pm0.08$ & $1.15\pm0.05$ & 0.41 & $7.49\pm0.26$ & $10\pm4$ \\
WF3/250 & 496 & 347 & $21.66\pm0.02$ & $-0.23\pm0.07$ & $0.49\pm0.04$ & $1.10\pm0.03$ & 0.47 & $7.59\pm0.18$ & $22\pm7$ \\
WF3/272 & 790 & 361 & $22.43\pm0.05$ & $-0.33\pm0.17$ & $0.59\pm0.10$ & $0.95\pm0.06$ & 0.93 & $7.33\pm0.35$ & $7\pm4$ \\
WF3/286 & 89 & 374 & $21.02\pm0.01$ & $-0.64\pm0.04$ & $0.30\pm0.03$ & $0.49\pm0.03$ & 0.22 & $<7.0$ & $13\pm3$ \\
WF3/321 & 451 & 400 & $21.98\pm0.03$ & $-0.06\pm0.16$ & $0.67\pm0.06$ & $1.07\pm0.04$ & 0.32 & $7.74\pm0.27$ & $19\pm9$ \\
WF3/344 & 667 & 417 & $21.32\pm0.02$ & $-0.19\pm0.06$ & $0.47\pm0.03$ & $0.91\pm0.02$ & 0.27 & $7.67\pm0.17$ & $33\pm9$ \\
WF3/416 & 71 & 469 & $21.18\pm0.01$ & $-0.36\pm0.04$ & $0.36\pm0.03$ & $0.93\pm0.02$ & 0.29 & $7.47\pm0.15$ & $29\pm7$ \\
WF3/425 & 438 & 481 & $21.81\pm0.03$ & $-0.29\pm0.09$ & $0.63\pm0.05$ & $1.04\pm0.03$ & 0.58 & $7.37\pm0.21$ & $13\pm5$ \\
WF3/486 & 293 & 523 & $21.41\pm0.02$ & $-0.29\pm0.06$ & $0.44\pm0.03$ & $0.97\pm0.02$ & 0.35 & $7.53\pm0.16$ & $25\pm7$ \\
WF3/488 & 428 & 524 & $22.41\pm0.04$ & $0.03\pm0.21$ & $0.51\pm0.08$ & $0.90\pm0.05$ & 0.62 & $8.01\pm0.35$ & $18\pm11$ \\
WF3/497 & 308 & 533 & $21.90\pm0.03$ & $-0.08\pm0.13$ & $0.68\pm0.05$ & $1.30\pm0.03$ & 0.49 & $7.69\pm0.27$ & $20\pm9$ \\
WF3/502 & 152 & 536 & $21.49\pm0.02$ & $0.68\pm0.43$ & $1.18\pm0.06$ & $1.72\pm0.02$ & 0.62 & $8.95\pm0.51$ & $206\pm180$ \\
WF3/553 & 694 & 566 & $22.45\pm0.04$ & $-0.23\pm0.17$ & $0.59\pm0.09$ & $1.19\pm0.05$ & 0.44 & $7.51\pm0.34$ & $9\pm5$ \\
WF3/612 & 423 & 598 & $21.04\pm0.01$ & $-0.06\pm0.06$ & $0.45\pm0.03$ & $0.87\pm0.02$ & 1.08 & $7.90\pm0.17$ & $57\pm17$ \\
WF3/613 & 55 & 599 & $21.69\pm0.03$ & $-0.52\pm0.06$ & $0.42\pm0.05$ & $0.92\pm0.03$ & 0.32 & $7.09\pm0.19$ & $8\pm2$ \\
WF3/620 & 133 & 605 & $21.99\pm0.04$ & $-0.76\pm0.07$ & $0.24\pm0.07$ & $0.50\pm0.07$ & 0.47 & $<7.0$ & $5\pm3$ \\
WF3/630 & 133 & 609 & $22.16\pm0.04$ & $-0.78\pm0.09$ & $0.34\pm0.09$ & $0.71\pm0.07$ & 0.22 & $<7.0$ & $4\pm3$ \\
WF3/737 & 273 & 664 & $22.00\pm0.03$ & $-0.41\pm0.10$ & $0.77\pm0.07$ & $1.07\pm0.04$ & 0.21 & $<7.0$ & $5\pm2$ \\
WF3/742 & 637 & 665 & $22.43\pm0.04$ & $-0.24\pm0.13$ & $0.39\pm0.07$ & $0.74\pm0.05$ & 0.37 & $7.65\pm0.28$ & $11\pm5$ \\
WF3/748 & 326 & 666 & $20.30\pm0.01$ & $-0.75\pm0.02$ & $0.32\pm0.02$ & $0.54\pm0.01$ & 0.73 & $<7.0$ & $25\pm6$ \\
WF3/754 & 202 & 669 & $21.56\pm0.02$ & $0.03\pm0.12$ & $0.55\pm0.05$ & $1.08\pm0.03$ & 1.61 & $7.97\pm0.26$ & $39\pm17$ \\
WF3/802 & 674 & 696 & $22.33\pm0.04$ & $0.04\pm0.20$ & $0.38\pm0.07$ & $0.84\pm0.05$ & 1.25 & $8.13\pm0.46$ & $23\pm18$ \\
WF3/928 & 90 & 785 & $22.33\pm0.04$ & $-0.36\pm0.11$ & $0.34\pm0.06$ & $1.27\pm0.04$ & 0.58 & $7.49\pm0.24$ & $10\pm4$ \\
WF3/930 & 410 & 785 & $21.61\pm0.02$ & $-0.38\pm0.06$ & $0.38\pm0.04$ & $0.84\pm0.03$ & 0.21 & $7.42\pm0.18$ & $18\pm5$ \\
WF3/931 & 756 & 785 & $22.25\pm0.09$ & $-0.22\pm0.16$ & $0.56\pm0.11$ & $0.80\pm0.09$ & 0.86 & $7.55\pm0.33$ & $12\pm6$ \\
WF4/19 & 275 & 64 & $19.85\pm0.01$ & $-0.43\pm0.02$ & $0.55\pm0.01$ & $0.72\pm0.01$ & 0.29 & $7.17\pm0.14$ & $56\pm13$ \\
WF4/36 & 637 & 77 & $21.16\pm0.03$ & $-0.59\pm0.06$ & $0.42\pm0.05$ & $0.60\pm0.04$ & 0.49 & $<7.0$ & $11\pm4$ \\
WF4/38 & 636 & 78 & $21.20\pm0.02$ & $-0.59\pm0.06$ & $0.44\pm0.05$ & $0.70\pm0.04$ & 0.49 & $<7.0$ & $11\pm4$ \\
WF4/150 & 793 & 162 & $22.42\pm0.07$ & $-0.23\pm0.19$ & $0.37\pm0.11$ & $0.87\pm0.10$ & 2.20 & $7.69\pm0.32$ & $12\pm6$ \\
WF4/178 & 462 & 181 & $21.52\pm0.02$ & $-0.39\pm0.05$ & $0.30\pm0.03$ & $0.65\pm0.03$ & 0.36 & $7.47\pm0.16$ & $21\pm5$ \\
WF4/233 & 576 & 225 & $20.66\pm0.01$ & $-0.23\pm0.03$ & $0.42\pm0.02$ & $0.87\pm0.01$ & 0.74 & $7.65\pm0.14$ & $60\pm14$ \\
WF4/264 & 527 & 246 & $22.46\pm0.04$ & $0.21\pm0.28$ & $0.59\pm0.08$ & $1.00\pm0.05$ & 0.58 & $8.29\pm0.64$ & $26\pm28$ \\
WF4/280 & 186 & 259 & $22.31\pm0.04$ & $0.39\pm0.31$ & $0.53\pm0.07$ & $0.94\pm0.05$ & 0.82 & $8.55\pm0.26$ & $45\pm20$ \\
WF4/296 & 641 & 269 & $20.39\pm0.01$ & $-0.58\pm0.02$ & $0.25\pm0.02$ & $0.46\pm0.02$ & 0.62 & $7.14\pm0.14$ & $31\pm7$ \\
WF4/301 & 713 & 272 & $21.13\pm0.02$ & $-0.03\pm0.06$ & $0.41\pm0.03$ & $0.77\pm0.03$ & 0.71 & $7.98\pm0.17$ & $58\pm17$ \\
WF4/332 & 350 & 294 & $21.92\pm0.03$ & $-0.55\pm0.14$ & $1.08\pm0.09$ & $1.39\pm0.04$ & 1.70 & $<7.0$ & $5\pm1$ \\
WF4/371 & 759 & 318 & $22.04\pm0.03$ & $-0.29\pm0.11$ & $0.47\pm0.06$ & $1.10\pm0.04$ & 1.16 & $7.50\pm0.24$ & $13\pm5$ \\
WF4/373 & 734 & 322 & $20.95\pm0.01$ & $0.20\pm0.11$ & $0.87\pm0.03$ & $1.49\pm0.02$ & 0.42 & $8.02\pm0.21$ & $73\pm26$ \\
WF4/460 & 769 & 394 & $22.24\pm0.04$ & $-0.17\pm0.14$ & $0.34\pm0.08$ & $0.90\pm0.06$ & 2.65 & $7.80\pm0.25$ & $16\pm7$ \\
WF4/497 & 790 & 423 & $21.90\pm0.03$ & $-0.25\pm0.11$ & $0.57\pm0.06$ & $1.12\pm0.04$ & 0.99 & $7.50\pm0.24$ & $15\pm6$ \\
WF4/530 & 659 & 445 & $22.28\pm0.03$ & $0.30\pm0.45$ & $1.06\pm0.10$ & $1.61\pm0.04$ & 0.62 & $9.89\pm0.13$ & $706\pm158$ \\
WF4/551 & 249 & 463 & $22.41\pm0.04$ & $-0.60\pm0.14$ & $0.69\pm0.10$ & $1.06\pm0.06$ & 0.77 & $<7.0$ & $3\pm4$ \\
WF4/554 & 619 & 468 & $22.10\pm0.03$ & $0.26\pm0.20$ & $0.44\pm0.06$ & $0.92\pm0.04$ & 0.95 & $8.48\pm0.20$ & $49\pm16$ \\
WF4/606 & 260 & 500 & $21.45\pm0.02$ & $-0.60\pm0.06$ & $0.47\pm0.04$ & $0.68\pm0.03$ & 0.41 & $<7.0$ & $8\pm3$ \\
WF4/614 & 262 & 505 & $22.05\pm0.04$ & $-0.53\pm0.10$ & $0.47\pm0.07$ & $0.64\pm0.06$ & 0.41 & $7.00\pm0.30$ & $5\pm2$ \\
WF4/657 & 137 & 536 & $19.88\pm0.01$ & $0.07\pm0.03$ & $0.60\pm0.01$ & $1.05\pm0.01$ & 1.24 & $8.00\pm0.14$ & $191\pm46$ \\
WF4/719 & 181 & 579 & $22.38\pm0.04$ & $-0.66\pm0.08$ & $0.22\pm0.07$ & $0.45\pm0.06$ & 0.98 & $<7.0$ & $3\pm1$ \\
WF4/747 & 624 & 600 & $22.41\pm0.04$ & $-0.73\pm0.12$ & $0.66\pm0.10$ & $1.88\pm0.05$ & 0.55 & $<7.0$ & $3\pm0$ \\
WF4/755 & 63 & 609 & $21.64\pm0.02$ & $0.02\pm0.21$ & $0.94\pm0.07$ & $1.29\pm0.03$ & 2.47 & $7.66\pm0.34$ & $24\pm14$ \\
WF4/758 & 222 & 611 & $21.80\pm0.03$ & $-0.57\pm0.06$ & $0.33\pm0.05$ & $0.72\pm0.04$ & 0.50 & $7.07\pm0.21$ & $7\pm2$ \\
WF4/791 & 588 & 641 & $21.27\pm0.02$ & $0.45\pm0.38$ & $1.37\pm0.06$ & $1.86\pm0.02$ & 0.83 & $9.89\pm6.97$ & $1790\pm21478$ \\
WF4/822 & 409 & 672 & $22.42\pm0.04$ & $-0.04\pm0.18$ & $0.41\pm0.08$ & $0.94\pm0.05$ & 0.31 & $7.96\pm0.36$ & $17\pm10$ \\
WF4/824 & 398 & 674 & $22.20\pm0.04$ & $-0.21\pm0.13$ & $0.57\pm0.07$ & $1.05\pm0.04$ & 0.30 & $7.56\pm0.28$ & $12\pm6$ \\
WF4/849 & 369 & 694 & $21.62\pm0.02$ & $-0.37\pm0.08$ & $0.49\pm0.05$ & $0.98\pm0.03$ & 2.77 & $7.34\pm0.20$ & $15\pm5$ \\
WF4/851 & 417 & 695 & $22.12\pm0.03$ & $-0.25\pm0.10$ & $0.42\pm0.06$ & $0.90\pm0.04$ & 0.58 & $7.61\pm0.23$ & $14\pm5$ \\
WF4/895 & 768 & 757 & $19.93\pm0.01$ & $-0.50\pm0.02$ & $0.41\pm0.01$ & $0.71\pm0.01$ & 0.45 & $7.15\pm0.14$ & $49\pm11$ \\
WF4/905 & 779 & 765 & $22.35\pm0.06$ & $-0.52\pm0.15$ & $0.57\pm0.11$ & $0.71\pm0.08$ & 0.51 & $<7.0$ & $3\pm3$ \\
WF4/908 & 792 & 769 & $22.09\pm0.06$ & $-0.64\pm0.12$ & $0.41\pm0.10$ & $0.50\pm0.10$ & 0.98 & $<7.0$ & $4\pm3$ \\
WF4/926 & 392 & 784 & $22.41\pm0.04$ & $-0.09\pm0.16$ & $0.28\pm0.07$ & $0.80\pm0.05$ & 1.00 & $7.98\pm0.32$ & $18\pm10$ \\
\enddata
\end{deluxetable}

\begin{deluxetable}{lccccccc}
\tablewidth{0pt}
\tabletypesize{\footnotesize}
\tablecaption{\label{tab:lffit}Luminosity function fits to cluster candidates.
$N$ is the number of clusters fitted. the column labeled $\alpha$ gives the 
exponent for fits of the form $dN(L)/dL = \beta \, L^{\alpha}$ while 
$a$ and $b$ refer to fits of the form 
$\log \scl[] [{\rm kpc}^{-2}] = b + a M_V$.  The last column lists 
$c = \beta/\ssfr$ for $\alpha = -2.4$. Milky Way numbers in paranthesis
are based on an assumed LF with $\alpha=-2.4$, normalized to the observed
cluster density near the Sun at $M_V=-8$.
}
\tablehead{Galaxy  & Fit interval &  $N$  &  $b$   & $a$ & $\alpha$ & $\scl[-8]$ & $c$ \\
            (1)    &     (2)      &  (3)  &  (4)   & (5) &   (6)    &    (7)     &   (8) }
\startdata
NGC628 & $-10.00 < M_V < -7.75$ & 60 & $3.96 \pm 0.95$ & $0.46 \pm 0.11$ & $-2.16 \pm 0.26$ & 1.8 & $7.96\times10^{-6}$ \\
NGC1313 & $-8.00 < M_V < -6.00$ & 259 & $3.70 \pm 0.35$ & $0.40 \pm 0.05$ & $-2.01 \pm 0.12$ & 2.9 & $5.94\times10^{-6}$ \\
NGC1313 & $-9.00 < M_V < -7.50$ & 52 & $5.27 \pm 1.58$ & $0.60 \pm 0.19$ & $-2.51 \pm 0.47$ & 2.7 & $5.58\times10^{-6}$ \\
NGC3184 & $-9.50 < M_V < -7.00$ & 72 & $4.56 \pm 0.88$ & $0.57 \pm 0.11$ & $-2.42 \pm 0.26$ & 1.0 & $4.80\times10^{-6}$ \\
NGC5236 & $-8.00 < M_V < -6.00$ & 228 & $4.70 \pm 0.34$ & $0.50 \pm 0.05$ & $-2.25 \pm 0.12$ & 5.1 & $3.03\times10^{-6}$ \\
NGC5236 & $-9.00 < M_V < -7.50$ & 34 & $5.23 \pm 2.24$ & $0.58 \pm 0.27$ & $-2.44 \pm 0.67$ & 4.2 & $2.52\times10^{-6}$ \\
NGC6744 & $-9.00 < M_V < -6.00$ & 60 & $3.55 \pm 0.36$ & $0.46 \pm 0.05$ & $-2.14 \pm 0.12$ & 0.8 & $1.05\times10^{-5}$ \\
NGC6946 & $-10.75 < M_V < -8.00$ & 243 & $5.46 \pm 0.45$ & $0.57 \pm 0.05$ & $-2.42 \pm 0.12$ & 8.1 & $1.45\times10^{-5}$ \\
LMC & $-9.50 < M_V < -6.00$ & 251 & $3.56 \pm 0.26$ & $0.40 \pm 0.03$ & $-2.01 \pm 0.08$ & 2.2 & $1.21\times10^{-5}$ \\
Milky Way  & -  &  -  & (4.43)   &  (0.56)  &   ($-2.4$)   &   0.89  & - \\
\enddata
\end{deluxetable}

\begin{deluxetable}{lccc}
\tablecaption{\label{tab:lmax}Galaxies from LR2000 and BHE02. Galaxies marked
with an asterisk ($\star$) are those studied in detail in the present paper. }
\tablehead{Galaxy  &   \ssfr   & $A$ & $M_V^{\rm max}$ (obs) \\
                   & $10^{-3} \, \msun \, {\rm kpc}^{-2}  \, {\rm yr}^{-1}$ &
		     kpc$^2$ & }
\startdata
\\
\multicolumn{4}{l}{From LR2000} \\
NGC45           & 0.23 & 110 & $-8.8$ \\
NGC247          & 0.18 & 200 & $-10.2$ \\
NGC300          & 0.49 & 159 & $-9.9$ \\
NGC628$^\star$  & 1.88 & 527 & $-11.3$ \\
NGC1156         & 3.07 & 60  & $-11.1$ \\
NGC1313$^\star$ & 4.04 & 105 & $-12.1$ \\
NGC2403         & 0.97 & 348 & $-9.9$ \\
NGC2835         & 0.73 & 126 & $-10.9$ \\
NGC2997         & 3.07 & 606 & $-12.9$ \\
NGC3184$^\star$ & 1.72 & 230 & $-10.6$ \\
NGC3621         & 1.67 & 527 & $-11.9$ \\
NGC4395         & 0.25 & 200 & $-9.1$ \\
NGC5204         & 0.83 & 38  & $-9.6$ \\
NGC5236$^\star$ & 13.76 & 166 & $-11.7$ \\
NGC5585         & 0.32 & 105 & $-10.8$ \\
NGC6744$^\star$ & 0.62 & 695 & $-11.0$ \\
NGC6946$^\star$ & 4.60 & 552 & $-13.0$ \\
NGC7424         & 0.18 & 960 & $-11.4$ \\
NGC7793         & 2.12 & 66 & $-10.4$ \\
\multicolumn{4}{l}{LR2000, litt.\ data} \\
NGC1569         & 9.43 & 13  & $-13.9$ \\
NGC1705         & 2.16 & 6   & $-13.7$ \\
NGC1741         & 12.78 & 382 & $-15.0$ \\
NGC5253         & 7.29 & 29  & $-11.1$ \\
IC1613          & 0.05 & 9 & $-5.8$ \\
LMC             & 1.52 & 79 & $-10.0$ \\
\multicolumn{4}{l}{From BHE02} \\
NGC4214         & 3.80 & 21 & $-12.04$ \\
NGC2366         & 1.95 & 19 & $-9.51$ \\
DDO50           & 1.26 &  8.6 & $-7.91$ \\
DDO168          & 0.85 & 5.1 & $-7.58$ \\
DDO165          & 0.18 & 3.0 & $-8.34$ \\
SexA            & 2.29  & 1.6 & $-7.12$ \\
\multicolumn{4}{l}{Lick data} \\
NGC3521         & 3.58  & 382 & $-11.5$ \\
NGC4258         & 0.702 &1414 & $-12.6$ \\
NGC5055         & 2.98  & 504 & $-11.4$ \\
NGC5194         & 8.21 & 578 & $-12.8$ \\
IC2574          & 0.17  & 121 & $-10.5$ \\
\enddata
\end{deluxetable}

\end{document}